\pdfoutput=1

\documentclass[11pt,twoside,a4paper,cmspaper,final,collab]{cms-tdr}
\def\svnVersion{274608}\def\cmsCernNo{2013-037}\def\cmsCernDate{\today}\def\cmsMessage{Published in Physical Review D as \href{http://dx.doi.org/10.1103/PhysRevD.91.012007}{\doi{10.1103/PhysRevD.91.012007.}}}

\begin{document}\cmsNoteHeader{EXO-12-038}

\hyphenation{had-ron-i-za-tion}
\hyphenation{cal-or-i-me-ter}
\hyphenation{de-vices}
\RCS$Revision: 274582 $
\RCS$HeadURL: svn+ssh://svn.cern.ch/reps/tdr2/papers/EXO-12-038/trunk/EXO-12-038.tex $
\RCS$Id: EXO-12-038.tex 274582 2015-01-22 03:12:35Z zuranski $

\newcommand{\Higgs}{\ensuremath{\PH}\xspace}
\newcommand{\Zprime}{\ensuremath{\cPZpr}\xspace}
\newcommand{\X}{\ensuremath{\cmsSymbolFace{X}}\xspace}
\newcommand{\qq}{\ensuremath{\cPq\cPaq}\xspace}
\newcommand{\qqprime}{\ensuremath{\cPq\cPaq'}}
\newcommand{\Kshort}{\ensuremath{\PKzS}\xspace}
\newcommand{\squark}{\ensuremath{\PSQ}\xspace}
\newcommand{\asquark}{\ensuremath{\PSQ^{*}}\xspace}
\newcommand{\sqasq}{\ensuremath{\squark\asquark+\squark\squark}\xspace}
\newcommand{\Lxy}{\ensuremath{L_{xy}}\xspace}
\newcommand{\Lxytrack}{\ensuremath{L_{xy}^{\text{track}}}\xspace}

\renewcommand{\chiz}{\ensuremath{\widetilde{\chi}^{0}_{1}}\xspace}

\newlength\cmsFigWidth
\ifthenelse{\boolean{cms@external}}{\setlength\cmsFigWidth{0.85\columnwidth}}{\setlength\cmsFigWidth{0.4\textwidth}}
\ifthenelse{\boolean{cms@external}}{\providecommand{\cmsLeft}{top}\xspace}{\providecommand{\cmsLeft}{left}\xspace}
\ifthenelse{\boolean{cms@external}}{\providecommand{\cmsRight}{bottom}\xspace}{\providecommand{\cmsRight}{right}\xspace}
\ifthenelse{\boolean{cms@external}}{\providecommand{\CL}{C.L.\xspace}}{\providecommand{\CL}{CL\xspace}}
\cmsNoteHeader{EXO-12-038}
\title{Search for long-lived neutral particles decaying to quark-antiquark pairs in proton-proton collisions at \texorpdfstring{$\sqrt{s} = 8\TeV$}{sqrt(s) = 8 TeV}}

\date{\today}

\abstract{
A search is performed for long-lived massive neutral particles decaying
 to quark-antiquark pairs. The experimental signature is a distinctive topology of a pair of jets, originating at a
secondary vertex.
Events were collected with the CMS detector at the CERN LHC in proton-proton collisions at a center-of-mass energy of 8\TeV. The data analyzed correspond to an integrated luminosity of 18.5\fbinv.
No significant excess
is observed above standard model expectations.
Upper limits at 95\% confidence level are set on the production cross section of a heavy neutral scalar particle,
 \Higgs, in the mass range of 200 to 1000\GeV, decaying promptly into a pair of long-lived neutral \X
 particles in the mass
range of 50 to 350\GeV, each in turn decaying into a quark-antiquark pair. For \X with mean proper
decay lengths
of 0.4 to 200\unit{cm},
 the upper limits are typically 0.5--200\unit{fb}.
The results are also interpreted in the context of an R-parity-violating supersymmetric model with long-lived neutralinos decaying into a quark-antiquark pair and a muon.
For pair production of squarks that promptly decay to
neutralinos with mean proper decay lengths of 2--40\unit{cm}, the upper limits on the cross section are typically 0.5--3\unit{fb}.
The above limits are the most stringent on these channels to date.
}

\hypersetup{%
pdfauthor={CMS Collaboration},%
pdftitle={Search for long-lived neutral particles decaying to quark-antiquark pairs in proton-proton collisions at sqrt(s) = 8 TeV},%
pdfsubject={CMS},%
pdfkeywords={CMS, physics}}

\maketitle

\section{Introduction}
\label{sec:Introduction}
This paper presents a search for massive, long-lived exotic particles, decaying into
quark-antiquark pairs (\qq), using data collected with the CMS detector at the CERN LHC.
Quarks fragment and hadronize into jets of particles.
We therefore search for events
containing a pair of jets originating from a common secondary
vertex that lies within the volume of the CMS tracker and is significantly displaced
from the colliding beams.
This topological signature has the potential to provide clear evidence for
physics beyond the standard model (SM).

A number of theories of new physics beyond the standard model predict
the existence of massive, long-lived particles, which could
manifest themselves through nonprompt decays to jets. Such scenarios arise, for example,
in various supersymmetric (SUSY) models, such as ``split SUSY''
\cite{Hewett:2004nw} or SUSY with very weak R-parity violation \cite{Barbier:2004ez}.
Similar signatures also occur in ``hidden valley'' models \cite{Han:2007ae}, and \Zprime models
with long-lived neutrinos~\cite{Basso:2008iv}.

We present search results in the
context of two specific models, so as to give a quantitative indication of the typical sensitivity.
In the first model, a long-lived, scalar, neutral
exotic particle, \X, decays to $\qq$. It is pair-produced in the decay of
a non-SM Higgs boson (\ie  $\Higgs\to
2\X$, $\X\to \qq$ \cite{Strassler:2006ri}), where the \Higgs boson is produced through gluon-gluon
fusion. In the second model, the long-lived particle is a neutralino \chiz, which decays into
two quarks and a muon through an R-parity violating coupling. The neutralinos are produced in events containing a pair
of squarks, where a squark can decay via the process $\sQua \to \cPq \chiz \to \cPq \cPq' \cPaq'' \Pgm$
\cite{Barbier:2004ez}.
Both models predict up to two displaced
dijet vertices per event within the volume of the CMS tracker.
The event selection is optimized for best sensitivity to the \Higgs model.
The same event selection is then applied to the neutralino model to
yield an additional interpretation of the search result.

The CDF and D0 collaborations have performed searches for metastable particles decaying to b-quark jets
using data collected at the Fermilab Tevatron at $\sqrt{s} = 1.96\TeV$
\cite{Aaltonen:2011rja, Abazov:2009ik}.
The ATLAS collaboration interpreted a search for displaced dijets, sensitive to decay lengths of 1--20\unit{m},
in terms of limits on the \Higgs model~\cite{ATLAS:2012av}. ATLAS also used results of a similar search, one
with a much smaller data set than the one considered in this paper, to place limits on the neutralino model~\cite{Aad:2011zb}.
Previous searches by the CMS collaboration for long-lived particles utilized high-ionization signals, large
time-of-flight measurements, nonpointing photons
or leptons, and decays inside the CMS hadron calorimeter~\cite{Chatrchyan:2013oca,Chatrchyan:2012ir,Chatrchyan:2012jna,Chatrchyan:2012dxa}.

\section{CMS detector}

The central feature of the CMS apparatus is a superconducting solenoid of 6\unit{m} internal diameter, providing a magnetic field of 3.8\unit{T}. Within the superconducting solenoid volume are a silicon pixel and strip tracker, a lead tungstate crystal electromagnetic calorimeter, and a brass and scintillator hadron calorimeter, each composed of a barrel and two endcap sections. Muons are measured in gas-ionization detectors embedded in the steel flux-return yoke outside the solenoid. Extensive forward calorimetry complements the coverage provided by the barrel and endcap detectors.
A more detailed description of the CMS detector, together with a definition of the coordinate system used and the relevant kinematic variables, can be found in~\cite{Chatrchyan:2008zzk}.

The tracker plays an essential role in the reconstruction of displaced vertices.
It comprises a large silicon strip tracker surrounding several layers of silicon pixel detectors.
In the central region in pseudorapidity ($\eta$), the pixel tracker consists of three coaxial barrel
layers at radii between 4.4\cm and 10.2\cm and the strip tracker consists of ten coaxial
barrel layers extending outwards to a radius of 110\cm.
Both detectors are completed by endcaps
at either end of the barrel. Each endcap consists of two disks in the pixel tracker, and three small
and nine large disks in the strip tracker. Together they extend the acceptance of the tracker up to
$\abs{\eta}<2.5$. The pixel tracker provides three-dimensional hit position measurements. The
strip tracker layers measure hit position in $r\phi$
in the barrel, or $z\phi$
in the endcaps.
A subset of strip tracker layers carry a second strip detector module, mounted back to back to the
first module and rotated by a stereo angle of 100\unit{mrad},
which provides a measurement of the third coordinate ($z$ in the
barrel, $r$ in the endcaps).
The initial track candidates (track seeds) are formed using only those layers that
 provide three-dimensional hit positions (pixel layers or strip layers with a stereo module).
The outermost stereo layer in the barrel region is located at a radius
of 50\cm. The track reconstruction algorithm
can therefore reconstruct displaced tracks from particles decaying up to
radii of ${\sim}50\cm$ from the beam line.
 The
performance of the track reconstruction algorithms has been studied in simulation and with data
\cite{Chatrchyan:2014fea}.

The global event reconstruction \cite{CMS-PAS-PFT-09-001,CMS-PAS-PFT-10-001} is designed to reconstruct and identify each particle in the event using an optimized combination of all subdetector information.
For each event, hadronic jets are clustered from these reconstructed particles with the infrared- and collinear-safe
 anti-\kt algorithm \cite{Cacciari:2008gp}
with a distance parameter $R$ of 0.5.
The jet momentum, determined as the vectorial sum of all particle momenta in the jet,
is adjusted with corrections derived from Monte Carlo (MC) simulations, test beam results,
and proton-proton collision data \cite{Chatrchyan:2011ds}. The corrections also account for the presence
of multiple collisions in the same or the adjacent bunch crossing (pileup interactions)
\cite{Cacciari:2007fd}.

\section{Online data selection}
\label{sec:DataAndMCSamples}

For this analysis, we use a sample of pp collision data at a center-of-mass energy of 8\TeV
corresponding to an integrated
luminosity of $18.5\pm0.5\fbinv$ \cite{CMS-PAS-LUM-13-001}.
The data were collected with a dedicated displaced-jet trigger.
At the trigger level, hadronic jets are
reconstructed using only the
energy deposits in the calorimeter towers. As a first step, \HT, defined as the scalar sum of the
transverse energy of all jets that have transverse momentum $\pt > 40\GeV$ and $\abs{\eta}<3$, is required to be above 300\GeV.
Then primary vertices are reconstructed, using tracks reconstructed solely with
 the pixel detector, and the vertex with the highest squared \pt sum of
its associated tracks is chosen as the primary event vertex.
Jets are considered if they have
$\pt > 60\GeV$ and $\abs{\eta}<2$.
To associate tracks to jets, the full tracking algorithm is applied to tracker hits in
 a cone of size $\Delta R<0.5$ around each jet direction, with $\Delta R=\sqrt{\smash[b]{(\Delta \eta) ^2 + (\Delta \phi)^2}}$.
The selection on the jet pseudorapidity ensures that all tracks fall within the
tracker acceptance $\abs{\eta}<2.5$.
For each reconstructed track,
an impact parameter is computed by measuring the shortest distance between the extrapolated
trajectory and the primary vertex. In order to accept an event at the trigger level, we demand that
at least two of the selected jets pass the following criteria:
\begin{itemize}
\item the jet has no more than two associated tracks with three-dimensional impact parameters smaller
than 300\mum;
\item no more than 15\% of the jet's total energy is carried by associated tracks with transverse
impact parameters smaller than 500\mum.
\end{itemize}

\section{Monte Carlo simulation samples}
\label{sec:MCSamples}

Signal MC samples are generated at leading order with
 \PYTHIA~6.426~\cite{PYTHIA}, using the CTEQ6L1 parton distribution functions
 \cite{Pumplin:2002vw}.
We simulate \Higgs~production through gluon fusion ($gg \to \Higgs$).  Subsequently,
 the \Higgs~is forced to decay to two long-lived, spin~0 exotic particles
($\Higgs \to 2\X$), each decaying into a quark-antiquark pair ($\X \to \qq$)
of any flavor except \ttbar with equal probability.
Samples with different combinations of \Higgs~masses ($m_{\Higgs} = 200$, 400, 1000\GeV ) and \X~boson masses
($m_{\X} = 50$, 150, 350\GeV) are generated. The lifetimes of
\X bosons are chosen to give a mean transverse decay length of approximately
3\cm, 30\cm, and 300\cm in the laboratory frame. For the neutralino model, we simulate squark pair production,
assuming that all squark flavors have the same mass, and the subsequent squarks decay to \chiz.
We use several combinations of squark and neutralino
masses: $(m_{\sQua},m_{\chiz}) = (350, 150)$, $(700, 150)$, $(700, 500)$, $(1000, 150)$,
$(1000, 500)$, $(1500, 150)$, and $(1500, 500)\GeV$. The R-parity
violating coupling $\lambda^{'}_{211}$~\cite{Barbier:2004ez} is set to a nonzero value and enables the decay of the
\chiz into a muon, an up quark, and a down quark.
The values of $\lambda^{'}_{211}$ are chosen to give a mean transverse
decay length of approximately 20\cm.

Background MC samples, produced with the same generator 
and parton distribution functions as the signal samples, 
comprise 35~million QCD multijet events with $\hat{p}_\text{T}$ between 80 and 800\GeV.
In this analysis, the background level is estimated from data and
the simulated background samples are only used to find appropriate
background discrimination
variables.

For all samples, the response of the CMS detector is simulated
in detail using \GEANTfour.9.4~\cite{GEANT4}. The samples are then processed through the trigger emulation and
event reconstruction chain of the CMS experiment. In addition, simulated
minimum bias events are overlaid with the primary collision
to model the pileup distribution from data. For the data used in this analysis, the
average number of pileup interactions was 21 per bunch crossing.

\section{Event reconstruction and preselection}
\label{sec:reco}

The offline primary vertex selection is analogous to the procedure employed in the trigger
(Sec.~\ref{sec:DataAndMCSamples}), except that the vertices used are obtained from fully reconstructed tracks.
The primary vertex is required to have at least
four associated tracks and to be displaced from the center of the detector
 by no more than 2\cm in the transverse plane and no more than
24\cm in $z$.
Using offline reconstructed jets, a requirement of $\HT >
325\GeV$ is applied, after which the corresponding trigger filter is $>$90\% efficient.
Furthermore, events produced by known instrumental effects are rejected.

The selection of jet candidates from secondary displaced vertices begins by searching for
at least two jets with $\pt>60\GeV$ and $\abs{\eta}<2$, similar to the trigger jet selection.
Tracks
with  $\pt>1\GeV$ are associated with jets by
requiring their momentum vectors (determined at the point of closest approach to the beam line)
 to have $\Delta R<0.5$ relative to the jet momentum vector.
Tracks may be associated with more than one jet.
 The set of associated tracks is divided
into ``prompt'' tracks, defined as those with transverse impact parameter value
 less than 500\mum, and ``displaced'' tracks, with higher transverse impact parameter. This
 requirement imposed for the displaced tracks
is large enough to exclude most b-hadron decay products.

The long-lived particle candidates are formed from all possible pairs of jets.
 The jets in the event are reconstructed with the anti-\kt
algorithm with a distance parameter of 0.5.
Therefore, if $\Delta R$ between the quarks from the \qq system
is below 0.5, they will not be reconstructed as two distinct jets.

The two sets of displaced tracks, corresponding to the two jets, are merged and fitted to a common secondary
vertex using an adaptive vertex fitter \cite{Waltenberger:1166320}.
The vertex fitting procedure down-weights tracks
that seem inconsistent with the fitted vertex position, based on their $\chi^2$ contribution to the vertex.
To include a track in the vertex, its weight is required to be at least 50\%.
This procedure reduces the bias caused by tracks incorrectly assigned to the vertex, \eg tracks originating from
pileup interactions.
The secondary vertex fit is required
to have a $\chi^2$ per degree of freedom less than 5.
The distance in the transverse plane between the secondary and the primary vertices, \Lxy, must be at
least eight times larger than its uncertainty.
We require that the secondary vertex includes at least one track from each of the two jets. This
requirement greatly reduces the background contribution from vertices due to nuclear interaction in the tracker material.
The nuclear interaction vertices are characterized by low invariant mass of the outgoing tracks,
making it unlikely that the outgoing tracks are associated with two distinct jets.
The invariant mass formed from all tracks associated with the vertex, assuming the
pion mass for each track,
must be larger than 4\GeV and the magnitude of the vector \pt sum of all tracks must be larger than 8\GeV.
Vertices can be misreconstructed when displaced tracks originating from different physical vertices accidentally cross.
To suppress such vertices, for each of the vertex tracks
we count the number of missing tracker measurements along the trajectory starting from the secondary vertex position until
 the first measurement is found. We require that the number of missing measurements per track, averaged
 over all the tracks associated with the displaced vertex, is less than 2.

\begin{figure}[htbp]
\centering
\includegraphics[width=0.42\textwidth]{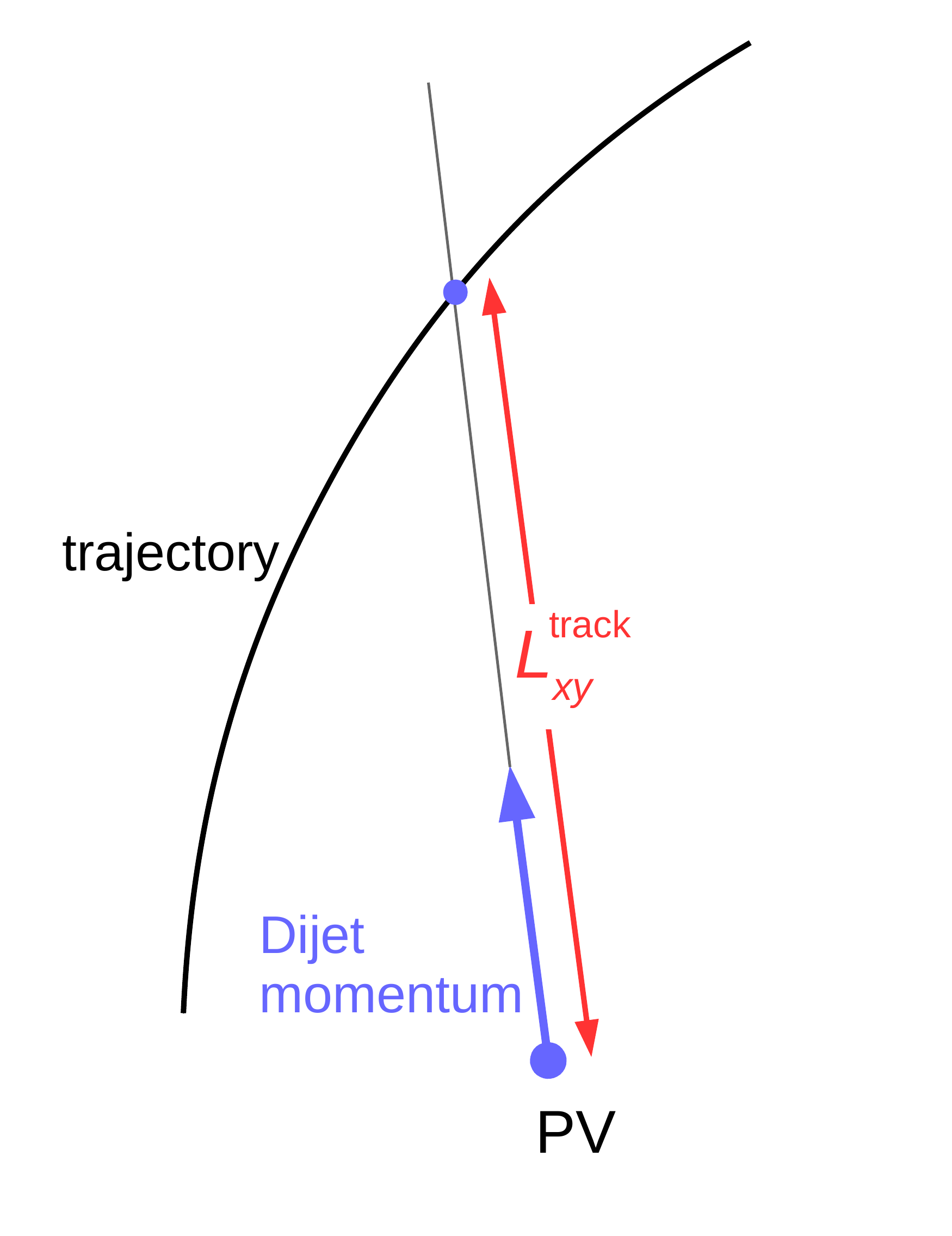}
\caption{
Diagram showing the calculation of 
the distance \Lxytrack.  In the transverse plane, 
\Lxytrack is the distance along the dijet momentum 
vector from the primary vertex (PV) to the point at which the track 
trajectory is crossed. \label{fig:guesslxy}}
\end{figure}

If a long-lived neutral particle decays into a dijet at a displaced location,
the trajectories of all tracks associated with the dijet should
cross the line drawn from the primary vertex in the direction of the dijet momentum vector
at the secondary vertex. The quantity \Lxytrack, illustrated in Fig.~\ref{fig:guesslxy},
is defined as the distance in the transverse plane between the primary vertex and the track trajectory, measured along the dijet
momentum direction. 
We use a clustering procedure to test whether the distribution of \Lxytrack is consistent with a displaced dijet hypothesis.
Clusters of maximum track multiplicity are obtained,
using a hierarchical clustering algorithm~\cite{opac-b1127878},
with a size
parameter which is set to 15\% of the distance \Lxy.
When multiple clusters are reconstructed,
we select the one whose mean \Lxytrack
 is closest to the value of \Lxy. For each dijet candidate, a reconstructed cluster with at least two tracks is required.

The candidate preselection, described above, may result in multiple dijet candidates per event.
The fraction of data events with more than one candidate passing the
preselection criteria is below 0.1\%. Nevertheless, for further event selection, we select
the best dijet candidate in each event, defined as the one with the highest track multiplicity for the
secondary vertex.

\section{Background estimation and final selection}
\label{sec:background}

The results are based on events for which the dijet candidate
passing the preselection criteria (Sec.~\ref{sec:reco}) also passes three additional selection
criteria.
For this purpose, the correlation factors
between the discriminating variables of the simulated background candidates have been studied, until a set
of three nearly independent criteria has been found.

\begin{figure*}[htbp]
\centering
\includegraphics[width=0.49\textwidth]{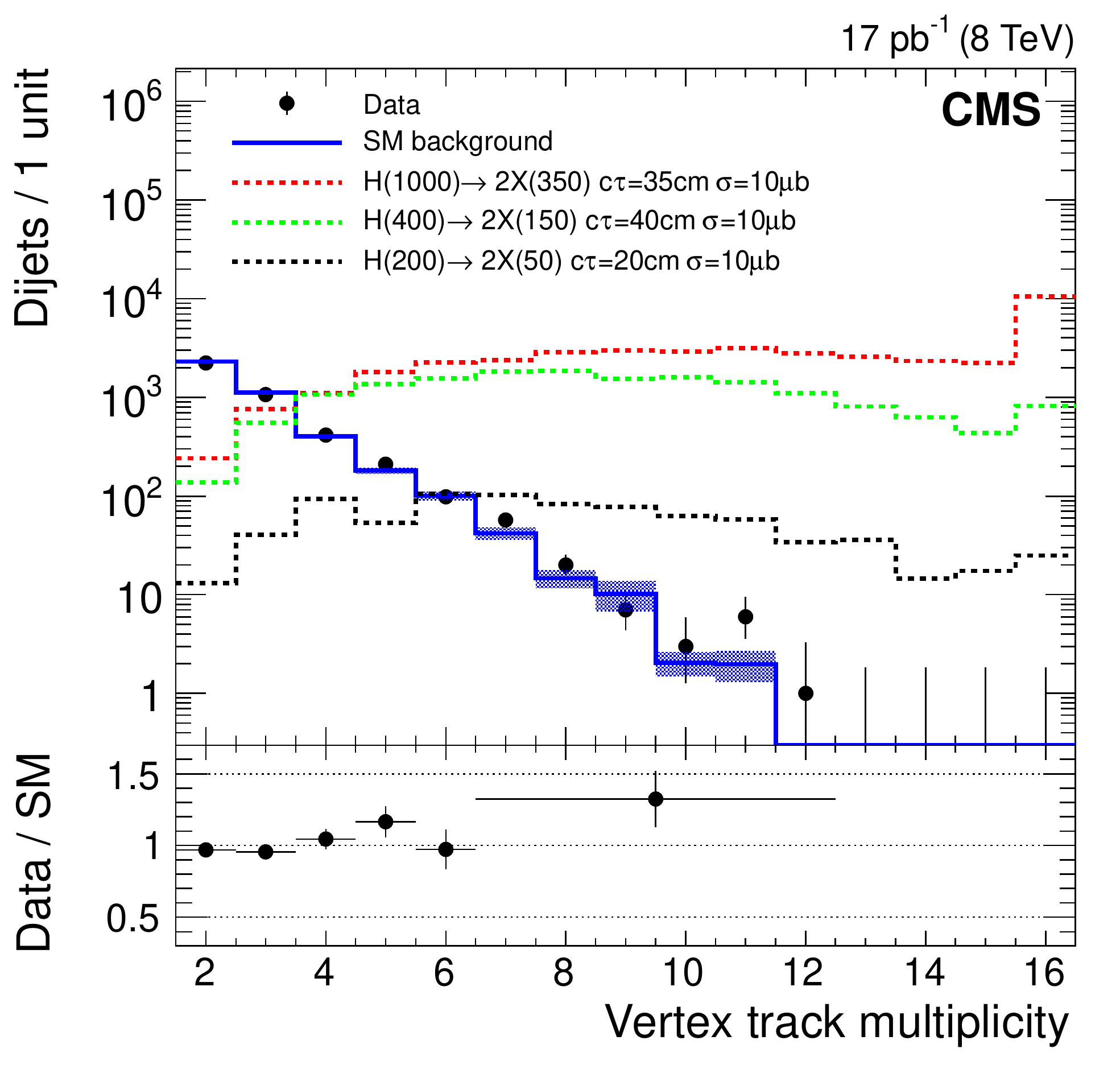}
\includegraphics[width=0.49\textwidth]{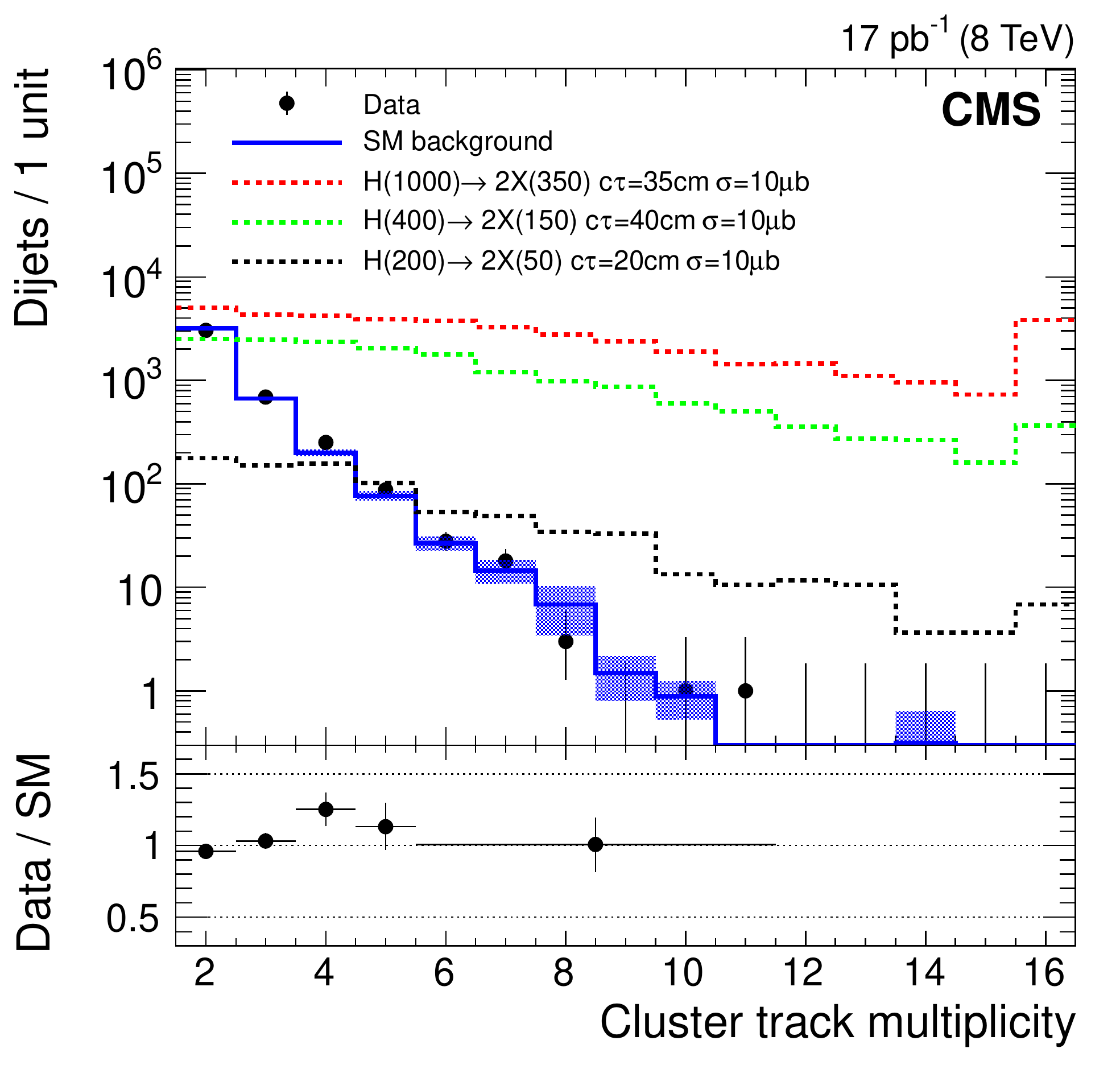}
\includegraphics[width=0.49\textwidth]{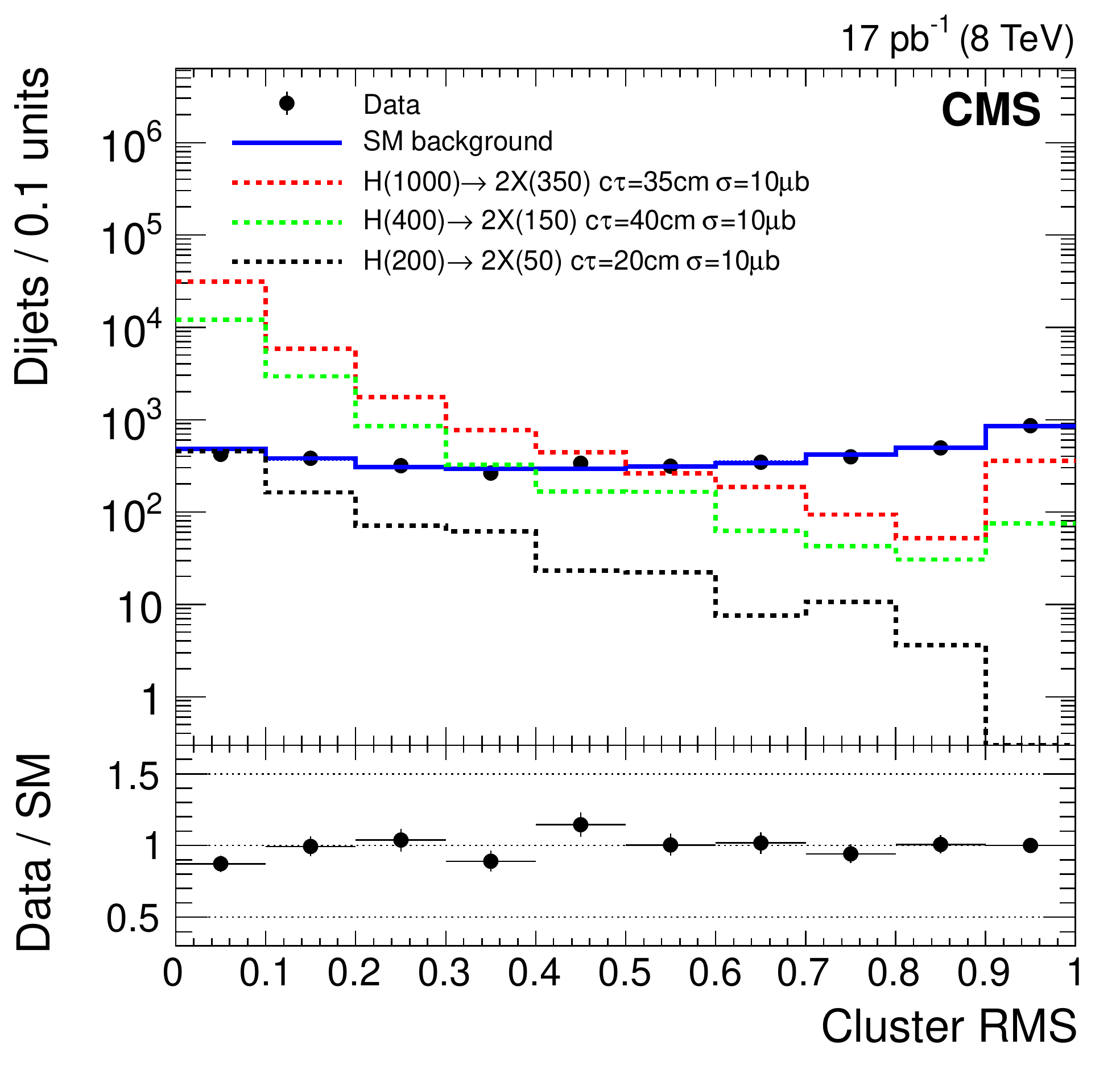}
\includegraphics[width=0.49\textwidth]{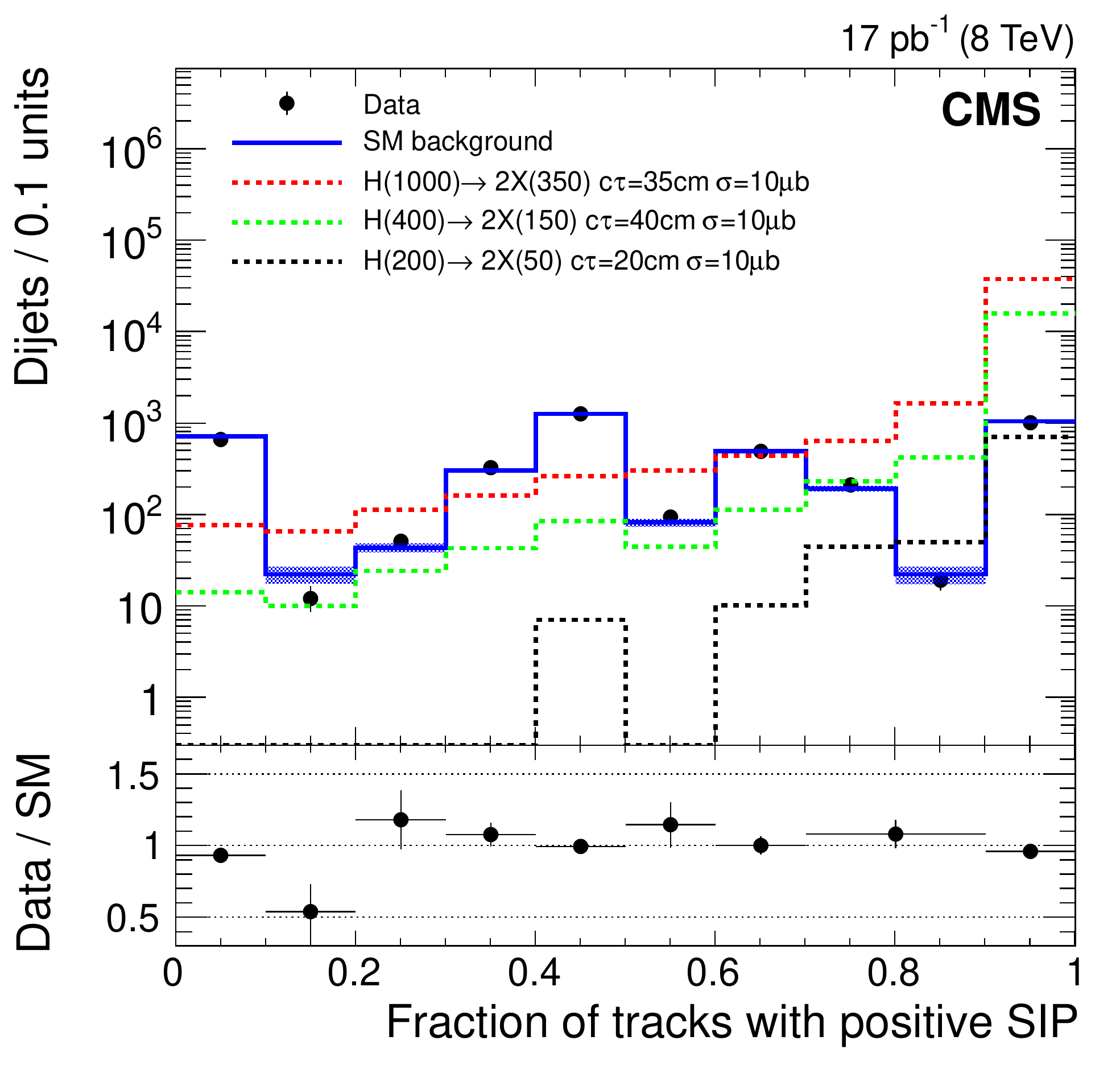}
\caption{Dijet variables employed in the likelihood discriminant for simulated signal, simulated SM
QCD background,
 and candidates in data, after the preselection. Data and simulated events are selected using a trigger that
requires $\HT>300\GeV$. The simulated signal and SM background distributions are scaled to an
integrated luminosity of 17\pbinv. For purposes of illustration, the signal process
$\Higgs \to 2\X \to 2\qq$ is assigned a 10\unit{$\mu$b} cross section for each mass pair.
The differences between the mass pairs arise mainly from differences in the kinematic acceptance.
Error bars and bands shown for the data, simulated SM background, and data/SM background ratio distributions, correspond to statistical uncertainties.
The data/SM ratio histograms are shown with neighboring bins merged, until the relative
statistical uncertainty is less than 25\%. The last bin in each histogram is an overflow bin.
\label{fig:disc}}
\end{figure*}

The first two selection criteria consist of simultaneous requirements on the number of prompt tracks
and on the jet energy fraction of the prompt tracks, applied independently for each jet in the displaced dijet pair.
The third criterion is a likelihood discriminant, formed from the following four
variables:
\begin{itemize}
\item secondary vertex track multiplicity;
\item cluster track multiplicity;
\item cluster root mean square (RMS)---the relative RMS of \Lxytrack with respect to the value of \Lxy
for the secondary vertex,
for the displaced tracks associated with the cluster;
\item fraction of the secondary vertex tracks having a positive value of the signed impact parameter~(SIP). SIP is defined as
a scalar product between the vector pointing from the primary vertex to the point of closest approach
of the trajectory to the beam line
(impact parameter vector) and the dijet momentum vector.
\end{itemize}
 The likelihood ratio $p$ for an \X boson candidate is defined by:
\begin{equation}
p=\frac{p_\mathrm{S}}{p_\mathrm{S}+p_\mathrm{B}},
\end{equation}
with
\begin{equation}
p_{\mathrm{S(B)}}=\prod_{i=1}^{4}  p_{\mathrm{S(B)},i},
\label{eqn:likelihood}
\end{equation}
where $p_{\mathrm{S(B)},i}$ is the signal (background) probability density function for the $i${th} input
variable. The probability density functions $p_{\mathrm{S(B)},i}$ are obtained using normalized
signal and background MC distributions of dijet candidates passing the preselection.
Because of the limited number of events in the background MC samples, we select the MC events with
a looser trigger than the signal trigger, only requiring $\HT>300\GeV$ with no additional
requirement of a displaced dijet candidate. The same
loose trigger was in operation during data collection. However, only a fraction of the events passing the trigger
was recorded, so that the effective integrated luminosity for this data sample amounts to 17\pbinv.
Figure~\ref{fig:disc} presents the distributions of all four variables entering the likelihood discriminant
for data, SM background MC simulation, and signal MC samples.
The signal model distributions are
found to have little dependence on the input masses and lifetimes, and therefore all the signal samples
are merged in creating the $p_{\mathrm{S},i}$ functions.

The three selection criteria (number of prompt tracks and prompt track
energy fraction of jet 1, number of prompt tracks and prompt track energy fraction of
jet 2, and vertex/cluster discriminant) classify the events into eight regions.
\begin{table}[htbp]
\centering
\topcaption{Naming convention for the regions used in the background estimation procedure, A--G, and the signal region, H. The ``$+$" sign
corresponds to a selection
being applied and the ``$-$" sign to a selection being inverted. \label{tab:regions}}
\begin{scotch}{cccc}
\ifthenelse{\boolean{cms@external}}{
  Region & Jet 1& Jet 2& Vertex/cluster\\
  & selection& selection& selection\\
}{
  Region & Jet 1 selection & Jet 2 selection & Vertex/cluster selection \\
}
 \hline
 A & $-$ & $-$ & $-$ \\
 B & $+$ & $-$ & $-$ \\
 C & $-$ & $+$ & $-$ \\
 D & $-$ & $-$ & $+$ \\
 E & $-$ & $+$ & $+$ \\
 F & $+$ & $-$ & $+$ \\
 G & $+$ & $+$ & $-$ \\
 H & $+$ & $+$ & $+$ \\

\end{scotch}
\end{table}
As listed in Table \ref{tab:regions}, the events in the A region fail all three criteria,
 events in the B, C, D regions fail two of them and
pass one, events in the E, F, G regions fail one and pass two other criteria, and events in the
signal region H pass all the criteria. As the selection criteria are mutually
independent in background
discrimination, the background level in the signal region H can be estimated using
different products
of event counts in the other regions, namely FG/B, EG/C, EF/D, DG/A, BE/A, CF/A and  BCD/A$^2$.
We use BCD/A$^2$ for the background prediction because it yields the smallest statistical uncertainty.
If the selection criteria are perfectly independent, all of the above products predict statistically consistent amounts
of background. However, the spread of the background predictions
may be larger due to systematic effects
(\eg residual interdependence of the variables). We therefore assign the largest difference
between BCD/A$^2$ and the other six products as a conservative
 systematic uncertainty in the background
prediction.

We determine the numerical values of the selection criteria by optimizing the expected limit for
the \Higgs signal model.
Various values of the \Higgs mass, the \X mass, and the \X lifetime are considered.
The selection variables
do not strongly depend on the particle masses. Therefore, the optimal selection criteria
vary only as a function of the
mean transverse decay length of the generated \X bosons, $\langle \Lxy \rangle$.
We use two sets of selection criteria,
depending on whether
$\langle \Lxy \rangle$ is below or above 20\cm.
The selection criteria are
detailed in Table \ref{tab:background}. For the neutralino model, the lower lifetime
selection is used for all signal samples.

\begin{table*}[htbp]
\centering
\topcaption{Optimized selection criteria, the number of observed events in data, and the background expectations with their statistical (first) and systematic (second) uncertainties.
The low $\langle \Lxy \rangle$ selection is optimized for signal models with $\langle \Lxy \rangle < 20\cm$,
while the high  $\langle \Lxy \rangle$ selection is optimized for signal models with higher $\langle \Lxy \rangle$. \label{tab:background}}
\begin{scotch}{rcc}
 & low $\langle \Lxy \rangle$ selection & high $\langle \Lxy \rangle$ selection \\
\hline
Number of prompt tracks for each jet & ${\leq}1$ & ${\leq}1$ \\
Prompt track energy fraction for each jet & ${<}0.15$ & ${<}0.09$ \\
Vertex/cluster discriminant & ${>}0.9$ & ${>}0.8$  \\
\hline
Data events & 2 & 1 \\
Expected background & $ 1.56\pm0.25\pm0.47$ & $1.13\pm0.15\pm0.50$ \\
\end{scotch}
\end{table*}

\begin{figure}[htb]
\centering
\includegraphics[width=0.49\textwidth]{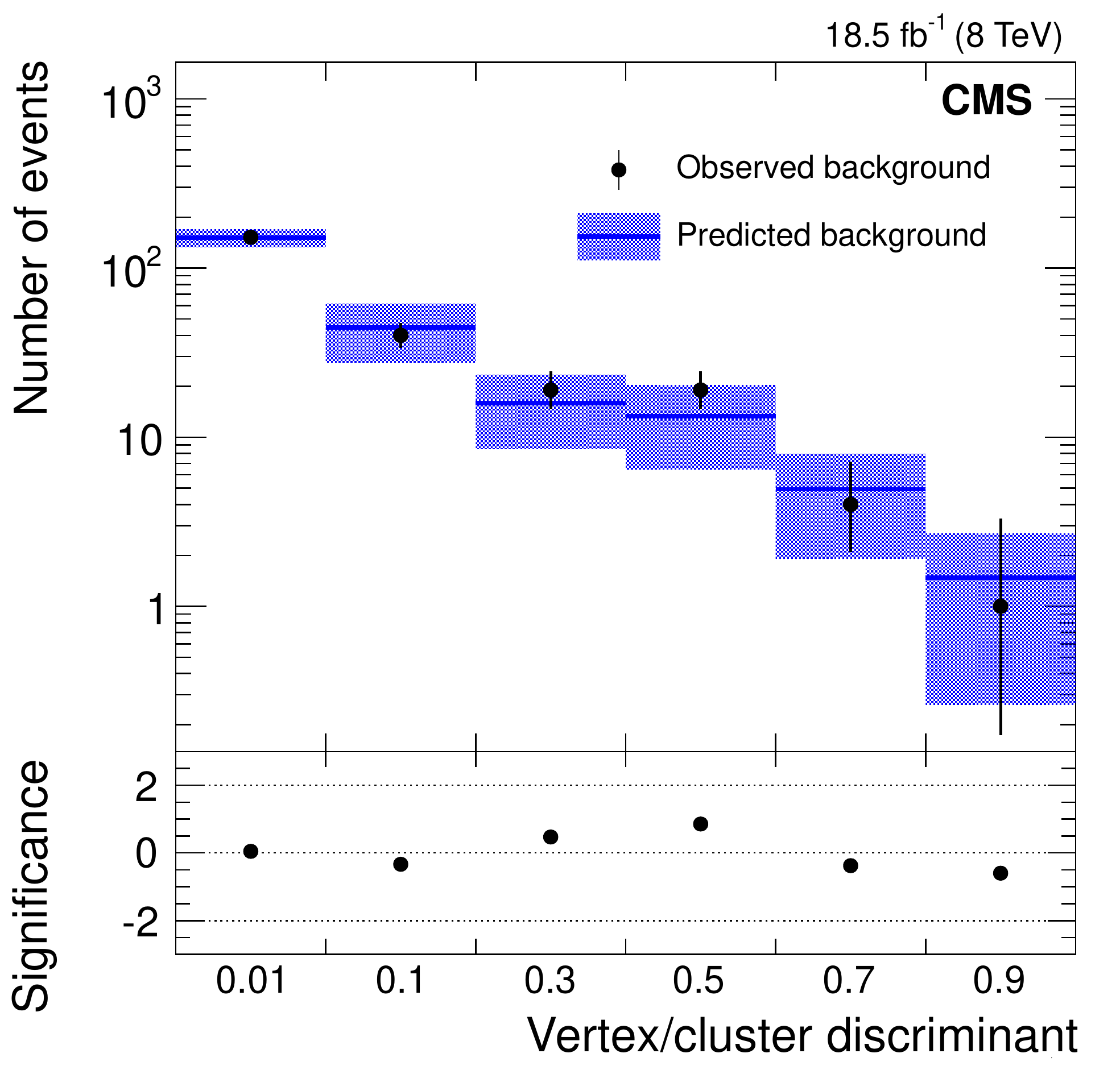}
\includegraphics[width=0.49\textwidth]{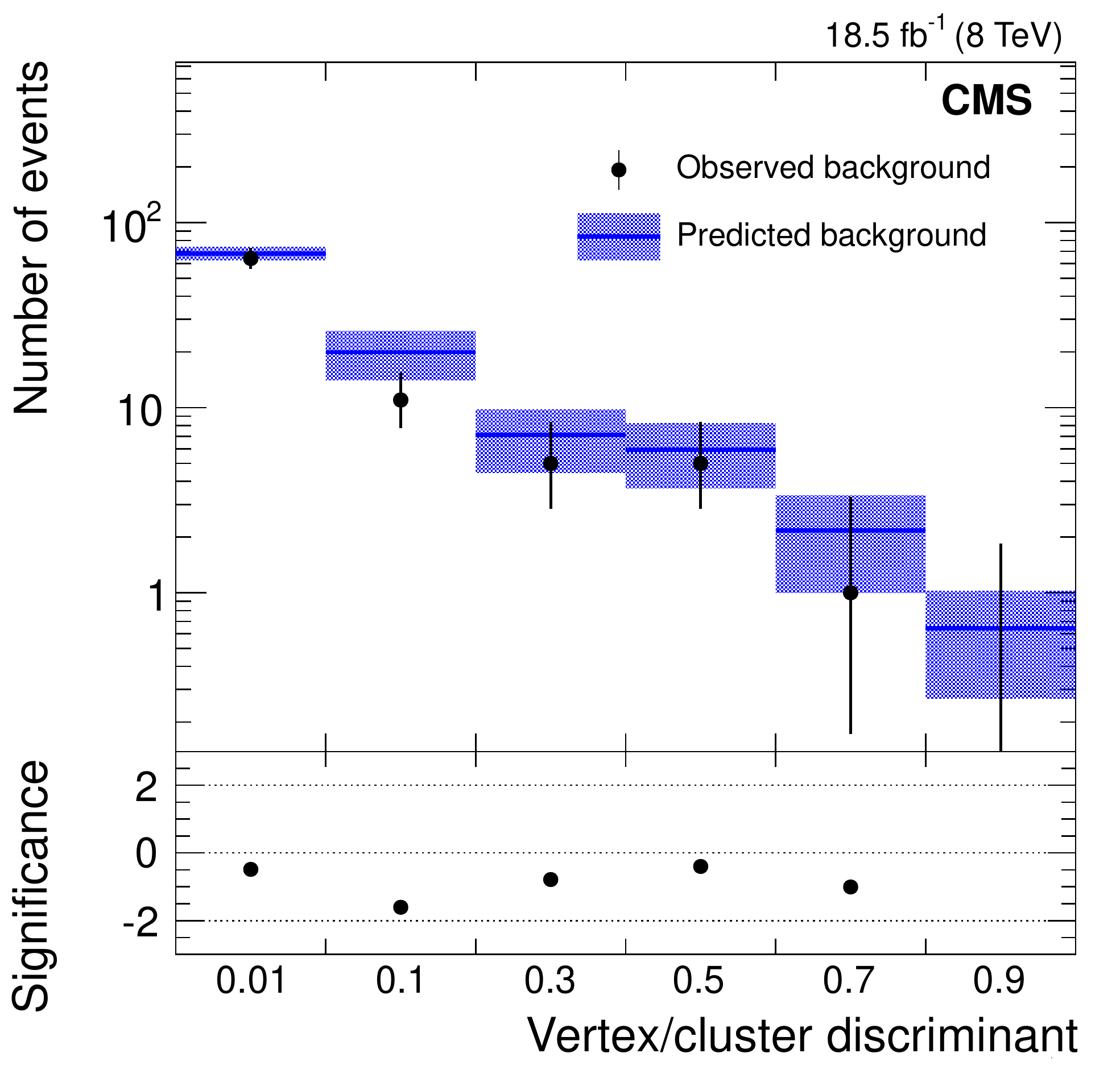}
\caption{
The expected and observed background level as a function of the vertex discriminant selection
 in the background dominated
data control region, obtained by inverting the selection requirement on missing track hits. The \cmsLeft\ (\cmsRight)
 plot is obtained after applying all other selection criteria as normal, optimized for the
region $\langle \Lxy \rangle <\,(>) 20\cm$. The predicted background error bands represent both statistical and systematic
uncertainties added in quadrature. \label{fig:bkg_NMiss}}
\end{figure}

To check the background prediction, a control region
is used that consists of events with a dijet candidate that is required
to pass all of the selection criteria but fail the preselection
requirement that the average number of missing measurements for dijet
tracks be less than 2.
 The signal efficiency in this region is a factor of 30 smaller than the efficiency
in the signal region, while the background level expectations are similar.
In Fig.~\ref{fig:bkg_NMiss}, we compare the observed background
as a function of the vertex discriminant in
this control sample, estimated using region H, against
the prediction from BCD/A$^2$.

We evaluate the $p$-value of the observed number of events based on a probability function that is a Poisson distribution convolved
with a Gaussian function representing the systematic uncertainty. In Fig.~\ref{fig:bkg_NMiss}, this $p$-value has been converted to an equivalent number of standard
deviations using the normal cumulative distribution. We refer to this number as the significance of the difference between the expected
and observed backgrounds. In all cases, the magnitude of the observed significance is less than 2 standard deviations.

\section{Systematic uncertainties}
\label{sec:systematics}

Sources of systematic uncertainty arise from the integrated luminosity,
background prediction, and signal efficiency estimation. The uncertainty in the integrated luminosity
measurement is 2.6\%~\cite{CMS-PAS-LUM-13-001}; the uncertainties in the background predictions are described in Sec.
\ref{sec:background}.

The signal efficiencies are obtained from MC simulations of the various signals, including full detector
response modeling. The systematic uncertainties related to the signal efficiency are dominated by the
differences between data and simulation, evaluated in control regions.
The relevant differences are discussed
below and their impact on the signal efficiency is evaluated. Table \ref{tab:signalsystematics}
summarizes
the sources of systematic uncertainty affecting the signal efficiency.

\begin{table}[htbp]
\topcaption{Systematic uncertainties affecting the signal efficiency.
For the uncertainties that depend on particle masses and lifetime,
a range of values is given for the signal parameters used.
In all cases, the uncertainties are relative.
\label{tab:signalsystematics}}
\centering
 \begin{scotch}{rl}

  Source & Uncertainty \\
  \hline
  Pileup modeling & 2\% \\
  Trigger efficiency & 6\% \\
  Jet energy corrections & 0\%--5\% \\
  Track finding efficiency & 4\% \\
  Jet momentum bias & 1\%--5\% \\
  \hline
 Total & 8\%--10\% \\

 \end{scotch}
\end{table}

Varying the modeling of the pileup, within its estimated uncertainty, yields a relative change in the signal
selection efficiency of less than 2\%, independent of masses and lifetimes over the ranges studied.

The trigger efficiency, obtained from control samples selected using lower threshold triggers, is found to be
higher in the simulation than in the data. An overall correction of $11\pm6\%$ is applied to the trigger efficiency.

Jet energy corrections are varied within their uncertainties \cite{Kirschenmann:2012toa}.
This variation affects only the \Higgs signal models
with $m_{\Higgs}=200$ and 400\GeV, with a relative change in the signal
 efficiency of 5\% and 3\%, respectively. For the \Higgs signal model with $m_{\Higgs}=1000$\GeV and for
 the neutralino model, the energies of the jets
are high enough that the variation in the energy correction
 does not alter the selection efficiency.

\subsection{Track finding efficiency}
\label{sec:trackEffi}

\begin{figure}[htb]
\centering
\includegraphics[width=0.49\textwidth]{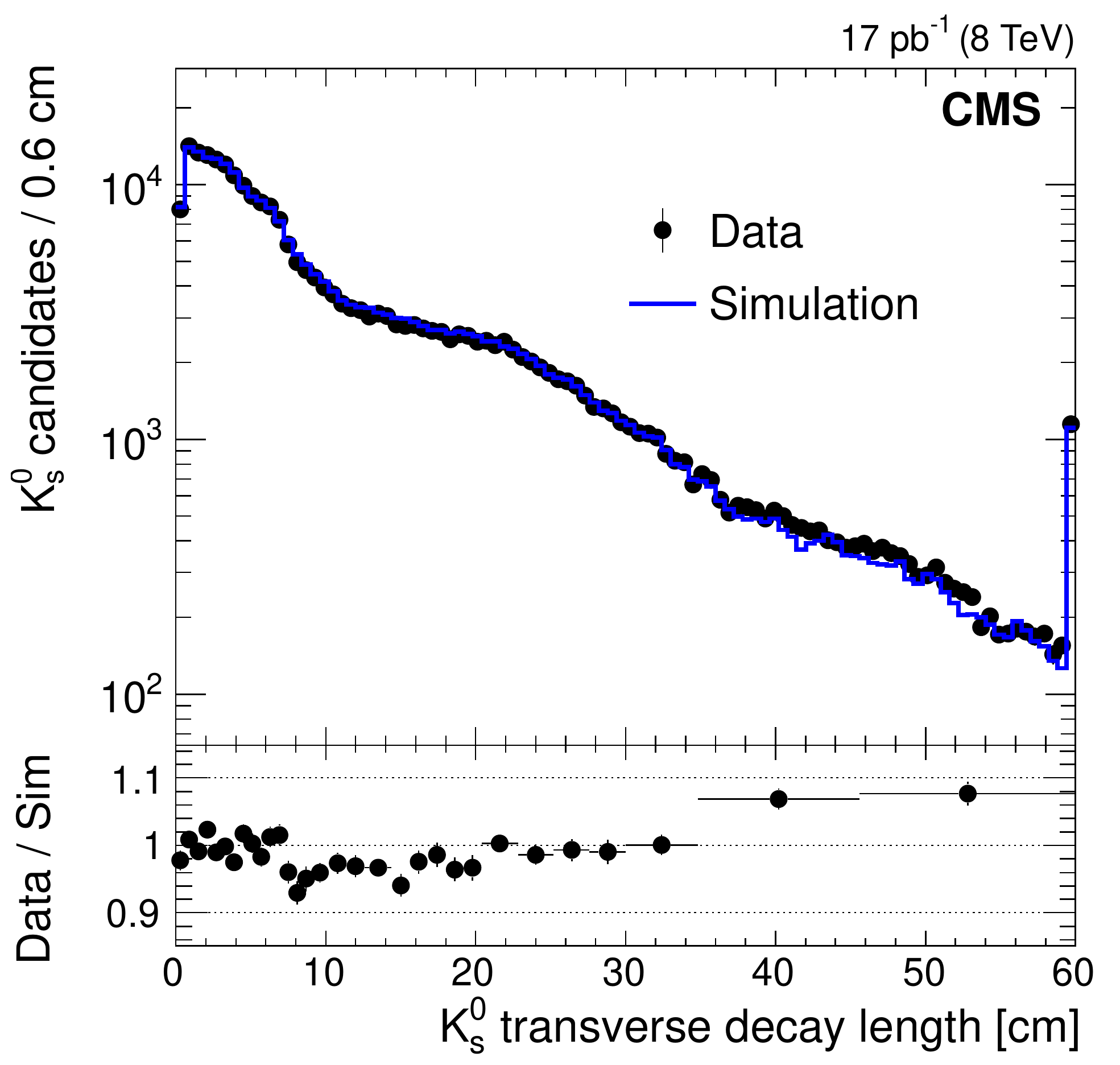}
\includegraphics[width=0.49\textwidth]{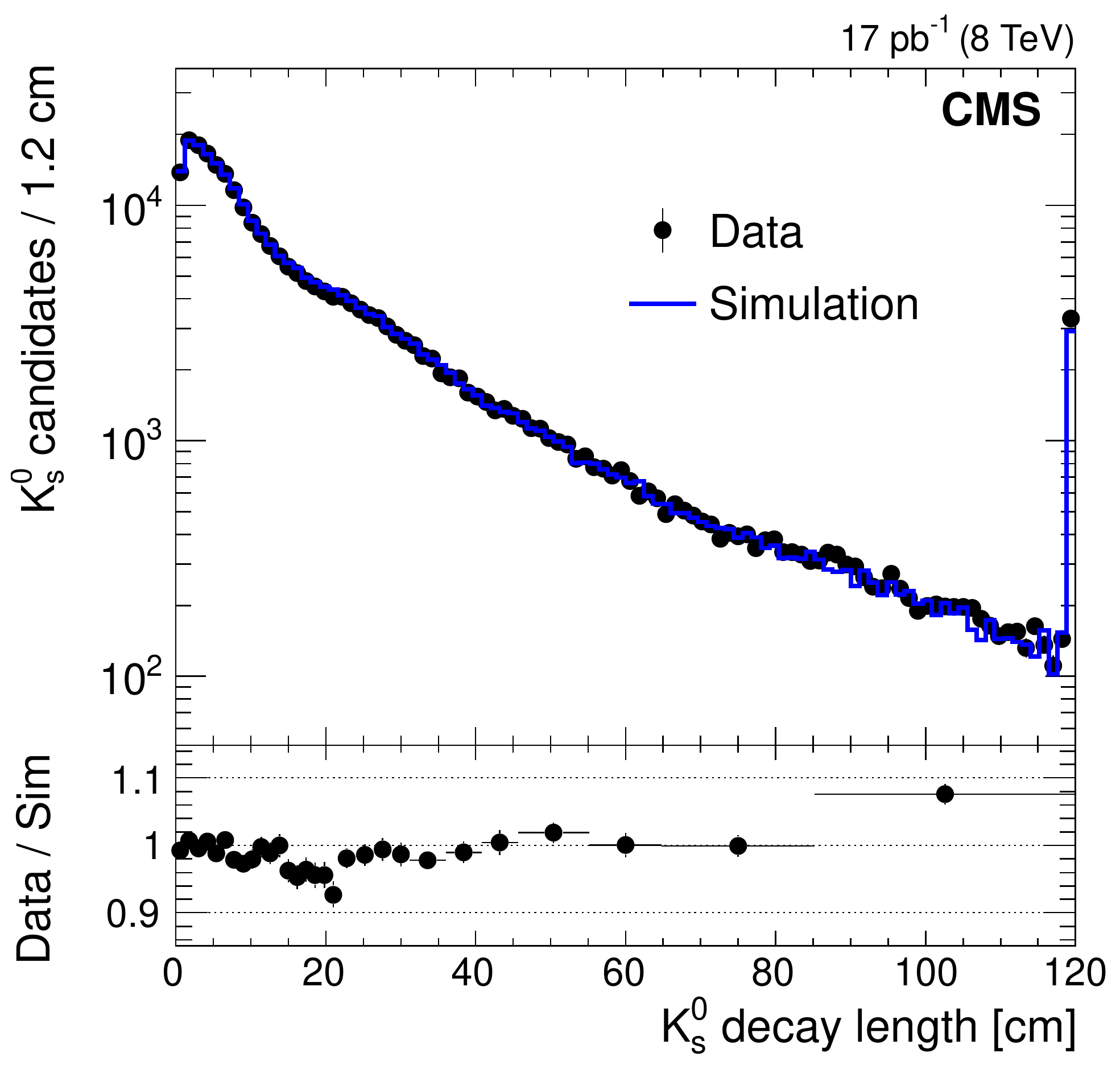}\\
\caption{Transverse decay length (\cmsLeft) and decay length (\cmsRight) distributions
of the $\text{K}^0_\text{S}$ candidates in data and simulation. The ratio histograms are shown with neighboring bins
merged until the relative statistical uncertainty falls below 2\%. The last bin contains all candidates that
are above the plotted range.\label{fig:ksdisplacement}}
\end{figure}

The tracks associated with the dijet candidates
 correspond mostly to light hadrons originating at a displaced
 location. The $\Kshort \to \Pgpp\Pgpm$ decay
 provides an abundant source of displaced tracks owing to the \Kshort mean proper decay length
of 2.68\cm \cite{Beringer:1900zz}.
The reconstruction of a \Kshort
candidate depends upon the reconstruction of the two pions. Therefore it is proportional to the
square of the efficiency for finding displaced tracks.
Approximately 250\,000 \Kshort candidates are obtained from a data
 sample collected with a multijet trigger. \Kshort candidates from simulation are obtained
using QCD multijet samples.
The MC simulation does not reproduce perfectly either the overall production rate
for \Kshort, or their kinematic distributions \cite{Khachatryan:2011tm}.
In order to account for these differences,
we first select \Kshort candidates with transverse decay lengths $\Lxy<2\cm$, where tracking efficiency
is high and well simulated.
We then match the \pt and $\eta$ distributions for these candidates and obtain
weights, binned in \pt and $\eta$, as well as an overall scale factor, that are
applied to all \Kshort candidates.
Figure \ref{fig:ksdisplacement} presents the
decay length distributions of the \Kshort candidates in data and simulation after this reweighting.
Data and simulation agree within 10\% in the entire range of the tracker
acceptance. Therefore, we estimate the tracking
 efficiency systematic uncertainty to be 5\%.
We study the track finding systematic uncertainty by removing 5\% of tracks before
 dijet reconstruction and selection. For all signal models, the signal reconstruction efficiency
is lowered by at most 4\%.

\subsection{Jet momentum bias}

For jets originating at a location that is significantly displaced
 from the event primary vertex, the reduced track reconstruction efficiency
and an inclined approach angle at the calorimeter face result in a systematic underestimation
of the jet momentum by up to 10\%, as determined from simulation. We assume that the detector
geometry is well reproduced in the MC simulation, and study only the jet momentum dependence
on the reconstruction efficiency of
displaced tracks. A 5\% variation in the jet energy fraction carried by tracks,
 corresponding to the systematic uncertainty in the track finding efficiency (Sec.~\ref{sec:trackEffi}),
leads to a change in the signal efficiency of 1\%--5\%, over the range of signal models considered.

\subsection{Effect of higher-order QCD corrections}

The signal reconstruction efficiency is sensitive to the jet energy
scale variations, for the \Higgs signal model with \Higgs masses of 200\GeV and 400\GeV.
Therefore, it is also sensitive to the modeling of
 the \Higgs \pt spectrum, which may be influenced by higher-order QCD corrections.
To study this effect,
we reweight the leading-order \PYTHIA \Higgs \pt spectrum from our signal samples to match the corresponding distribution,
determined at next-to-leading order (NLO) using \POWHEG \cite{Nason:2004rx,Frixione:2007vw,Alioli:2010xd}.
For signal with masses $m_{\Higgs}= 200\,(400)$\GeV and
$m_{\X}=50\,(150)$\GeV, this reweighting increases the efficiency by 20\,(3)\%, while for
other \Higgs masses the effect is below 1\%.
Since the \Higgs signature simply relates to a benchmark model,
 we do not incorporate this variation as an additional systematic uncertainty.

\section{Results}
\label{sec:results}

No significant excess of events is observed over the predicted backgrounds. Two events pass the low
$\langle \Lxy \rangle$ selection $(\langle \Lxy \rangle < 20\unit{cm})$.
One of the two events passing the low $\langle \Lxy \rangle$ selection additionally passes
the high $\langle \Lxy \rangle$ selection $(\langle \Lxy \rangle > 20\unit{cm})$.
No additional candidates
pass the high $\langle \Lxy \rangle$ selection. Both of these results are in
 agreement with the background expectations quoted in Table~\ref{tab:background}.

\subsection{Upper limits}
\label{subsec:limits}

We set 95\% confidence level
(\CL) upper limits on the signal cross section for a counting experiment
using the CL$_\mathrm{s}$ method \cite{Read:2002hq, Junk:1999kv}. The limit calculation
takes into account the systematic uncertainties described in Sec.~\ref{sec:systematics} by introducing
a nuisance parameter for each uncertainty, marginalized by a log-normal prior distribution.

Upper limits are placed on the mean number of signal events $N_S$ that could pass
the selection requirements. The resulting observed upper limits on $N_S$ are 4.6 events
for the low $\langle \Lxy \rangle$ selection and 3.7 events for the high $\langle \Lxy \rangle$ selection.
 These limits are independent
of the particular model assumed for production of long-lived particles.

In addition, upper limits on the production cross section for the \Higgs and neutralino
models are determined. The efficiency of the full set of selection criteria for both signal models,
at all considered masses,
is presented in Table~\ref{tab:sigeff}.
\begin{table}[!htbp]
\centering
\topcaption{
Signal reconstruction efficiency $\epsilon$ for the \Higgs and neutralino models
in simulated signal samples.
The trigger and reconstruction efficiencies are both taken into account.
The uncertainties are statistical only.\label{tab:sigeff}}
\begin{scotch}{llccr@{$\,\pm\,$}l}
$m_{\Higgs}$\,[\GeVns{}] & $m_{\X}$\,[\GeVns{}] & c$\tau$\,[cm] & $\langle \Lxy \rangle$\,[cm] & \multicolumn{2}{c}{$\epsilon$ [\%]}\\
\hline
200 & 50 & 2 & 3 & 0.25 & 0.05 \\
200 & 50 & 20 & 30 & 0.15 & 0.04 \\[0.65ex]

400 & 50 & 0.8 & 2.6 & 5.6 & 0.2 \\
400 & 50 & 8 & 26 &  3.3 & 0.2 \\
400 & 50 & 80 & 260 & 0.3 & 0.1 \\[0.65ex]

400 & 150 & 4 & 3 & 15.6 & 0.4 \\
400 & 150 & 40 & 30 & 7.6 & 0.3 \\
400 & 150 & 400 & 300 & 0.6 & 0.1 \\[0.65ex]

1000 & 150 & 1 & 2.5 & 41.3 & 0.5 \\
1000 & 150 & 10 & 25 & 31.1 & 0.5 \\
1000 & 150 & 100 & 250 & 4.8 & 0.2 \\[0.65ex]

1000 & 350 & 3.5 & 2.9 & 49.2 & 0.5 \\
1000 & 350 & 35 & 29 & 30.9 & 0.5 \\
1000 & 350 & 350 & 290 & 4.4 & 0.2 \\
\hline
$m_{\sQua}$\,[\GeVns{}] & $m_{\chiz}$\,[\GeVns{}] & c$\tau$\,[cm] & $\langle \Lxy \rangle$\,[cm] & \multicolumn{2}{c}{$\epsilon$\,[\%]}\\
\hline
350 & 150 & 17.8 & 22 & 7.2 & 0.3 \\
700 & 150 & 8.1 & 20 & 13.6 & 0.3 \\
700 & 500 & 27.9 & 20 & 22.8 & 0.3 \\
1000 & 150 & 5.9 & 19 & 13.0 & 0.3 \\
1000 & 500 & 22.7 & 21 & 26.4 & 0.3 \\
1500 & 150 & 4.5 & 21 & 8.6 & 0.2 \\
1500 & 500 & 17.3 & 23 & 28.8 & 0.4 \\
\end{scotch}
\end{table}

\begin{figure*}[htbp]
\centering
\includegraphics[width=0.49\textwidth]{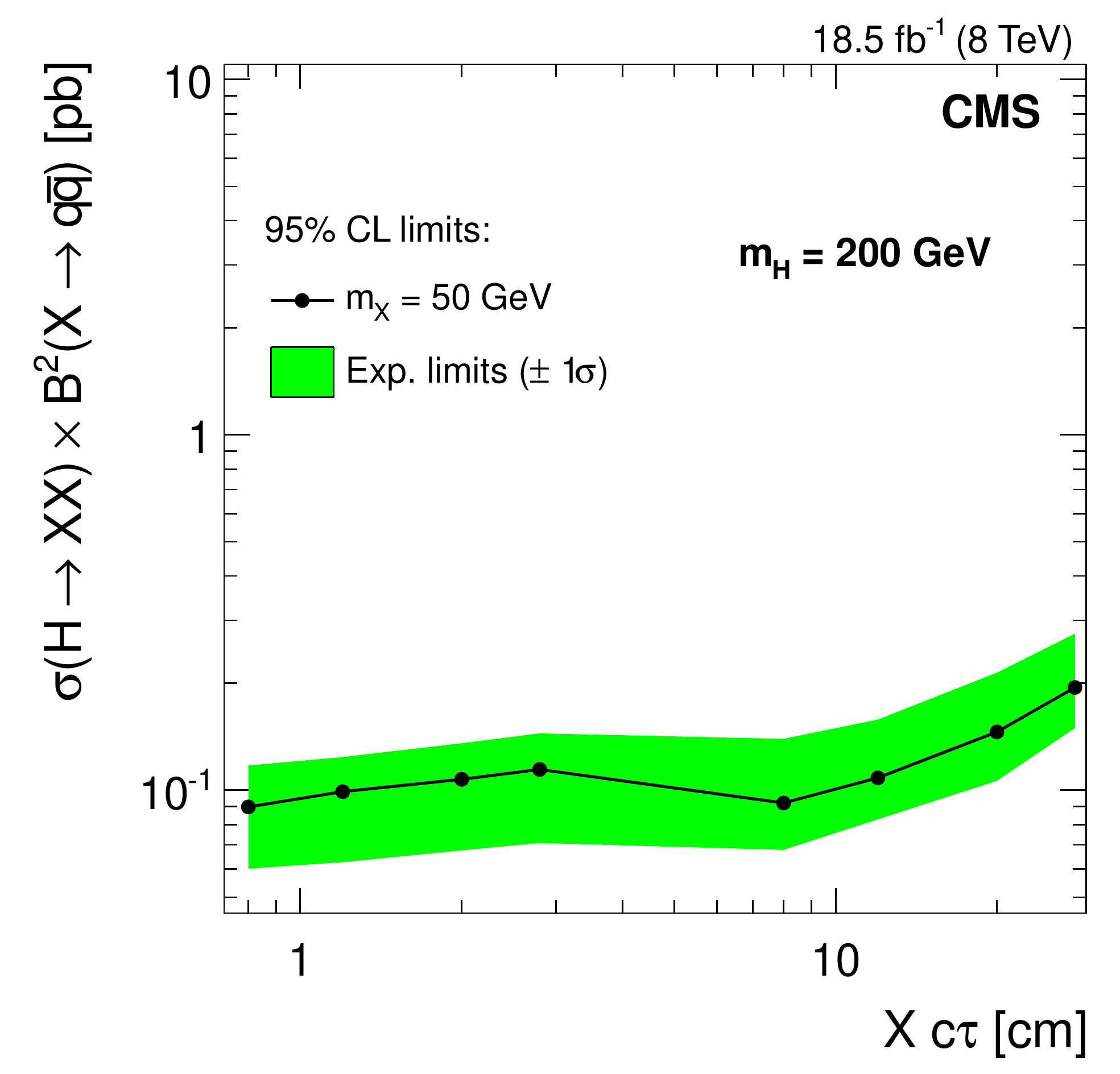}
\includegraphics[width=0.49\textwidth]{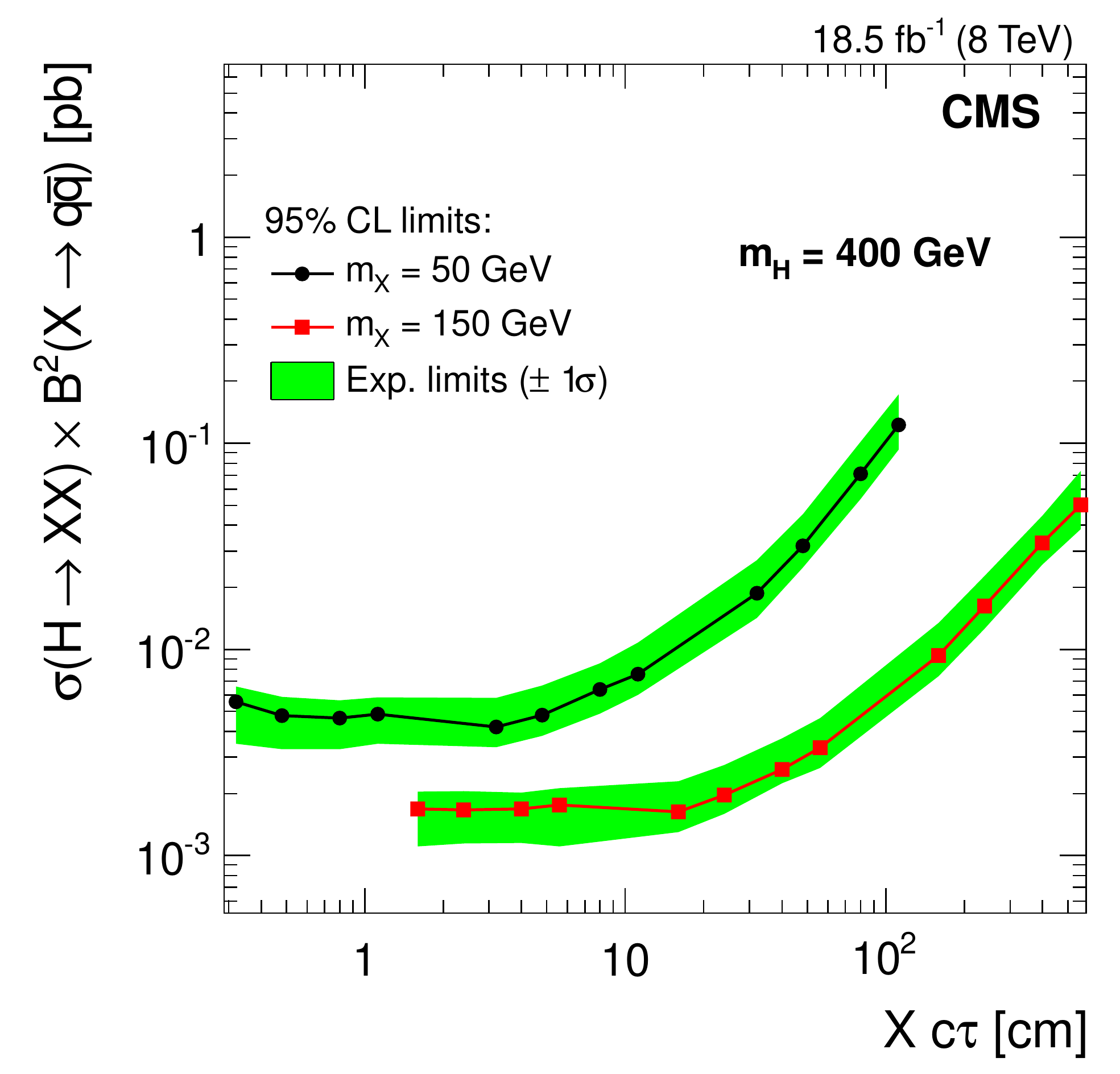}
\includegraphics[width=0.49\textwidth]{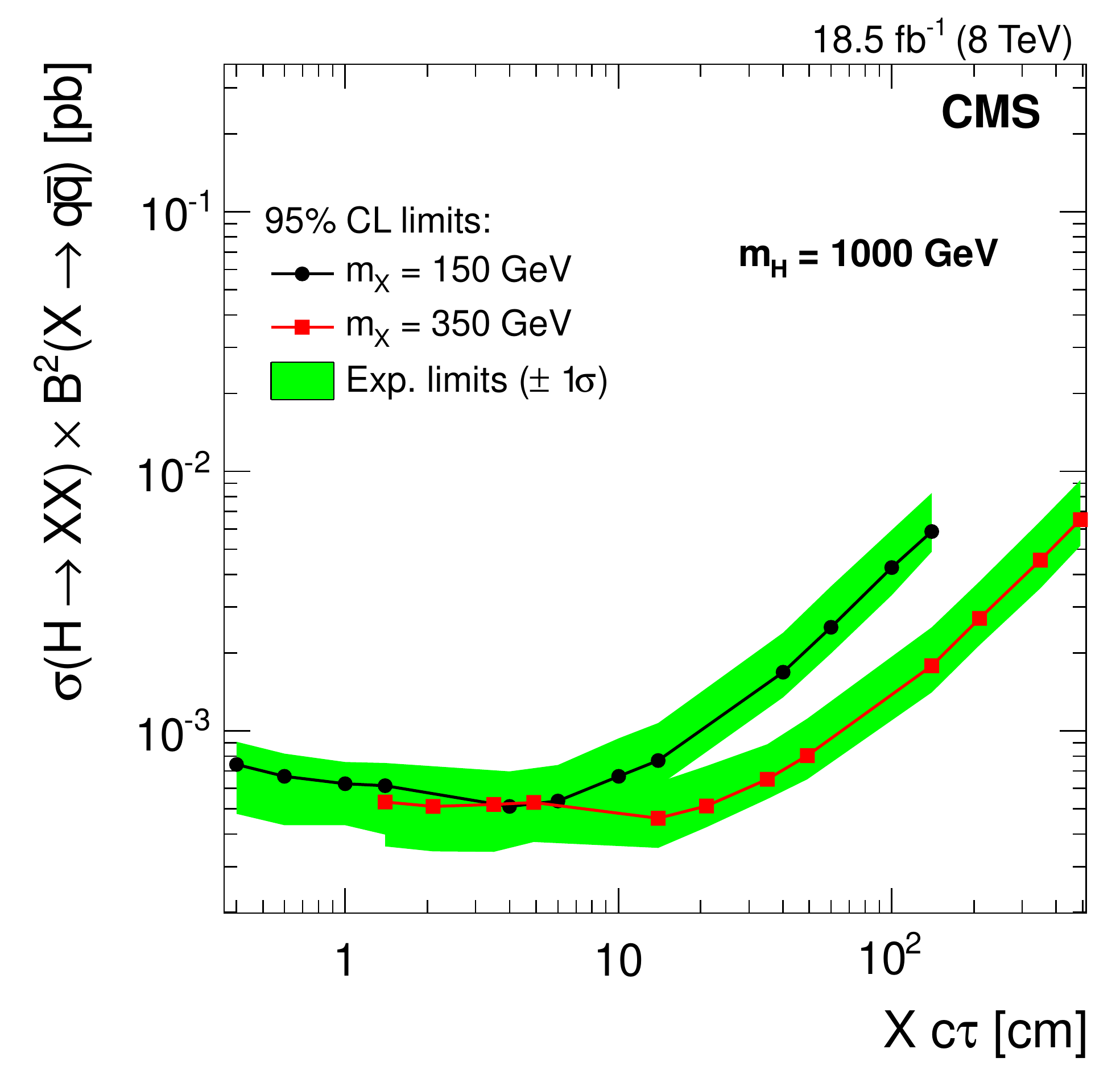}

\caption{The 95\% \CL upper limits on the product of the
cross section to produce a heavy resonance \Higgs
that decays to a pair of neutral long-lived particles
\X, and the branching fraction squared $\mathcal{B}^{2}$ for the \X decay into a quark-antiquark pair.
The limits are presented as a function of the \X particle mean proper decay length
separately for each \Higgs/\X mass point. Solid bands show the ${\pm}1\sigma$ range of variation of
the expected 95\% \CL limits.
\label{fig:limits}}
\end{figure*}

\begin{figure*}[htb]
\centering
\includegraphics[width=0.49\textwidth]{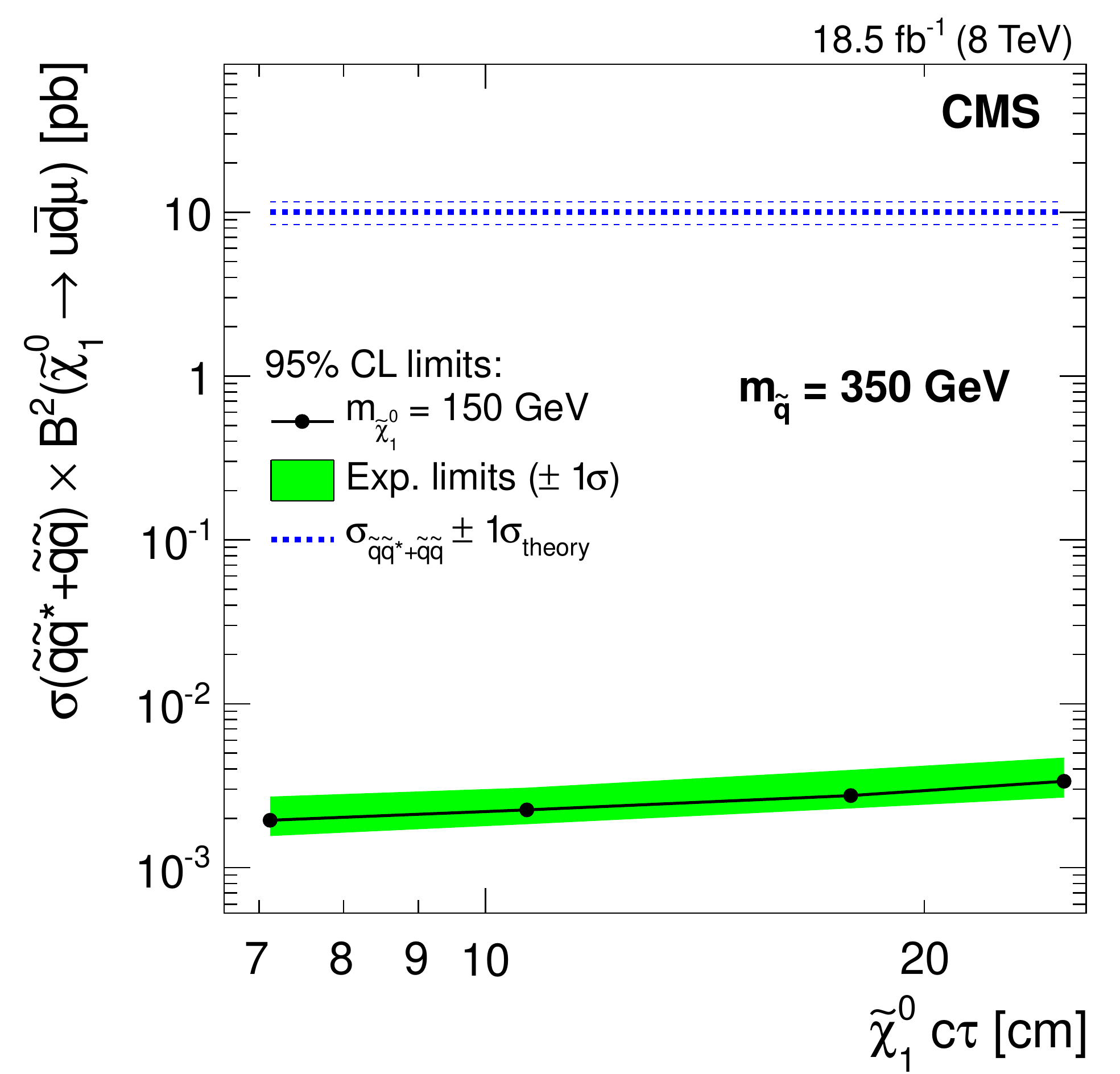}
\includegraphics[width=0.49\textwidth]{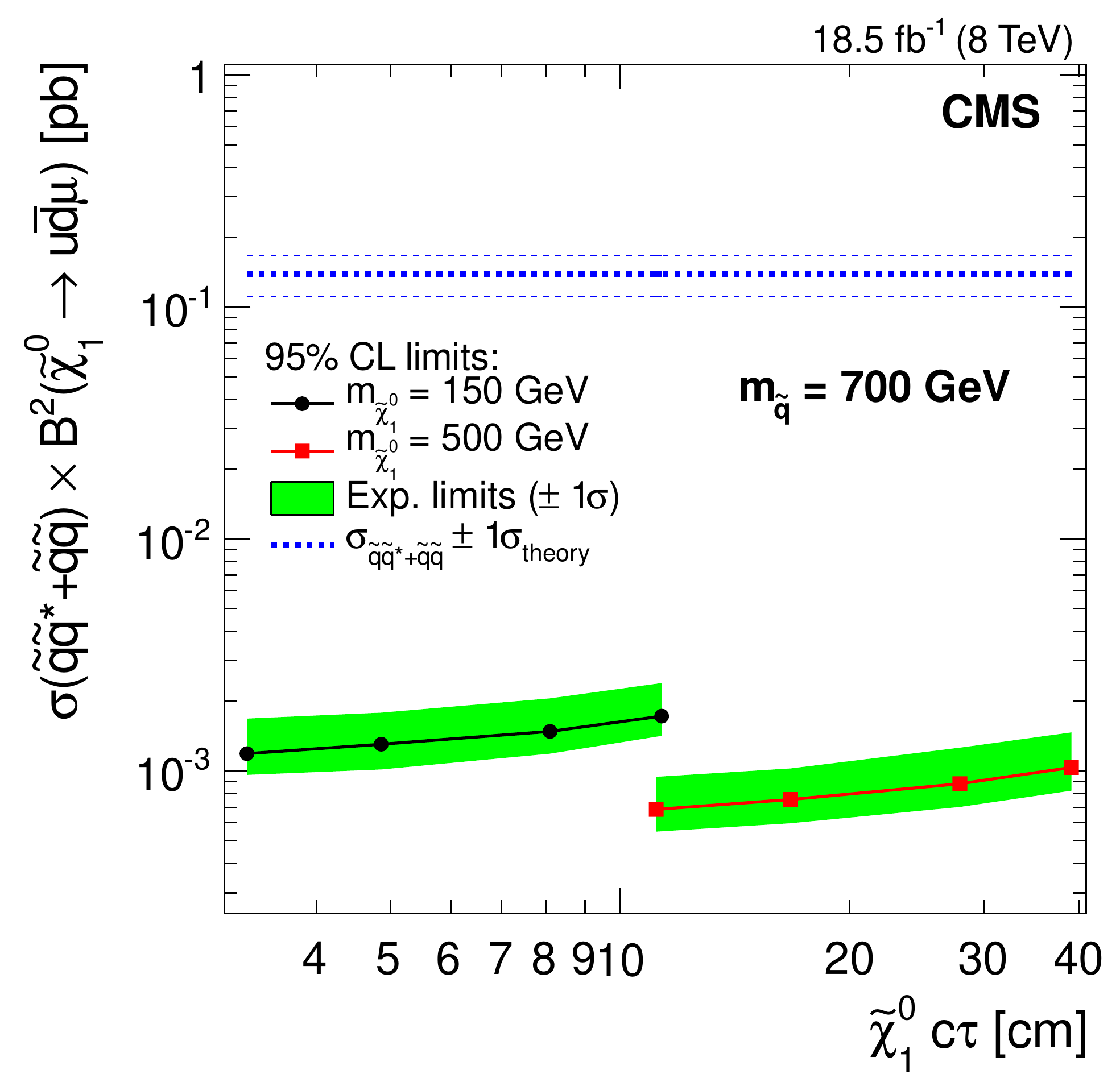}
\includegraphics[width=0.49\textwidth]{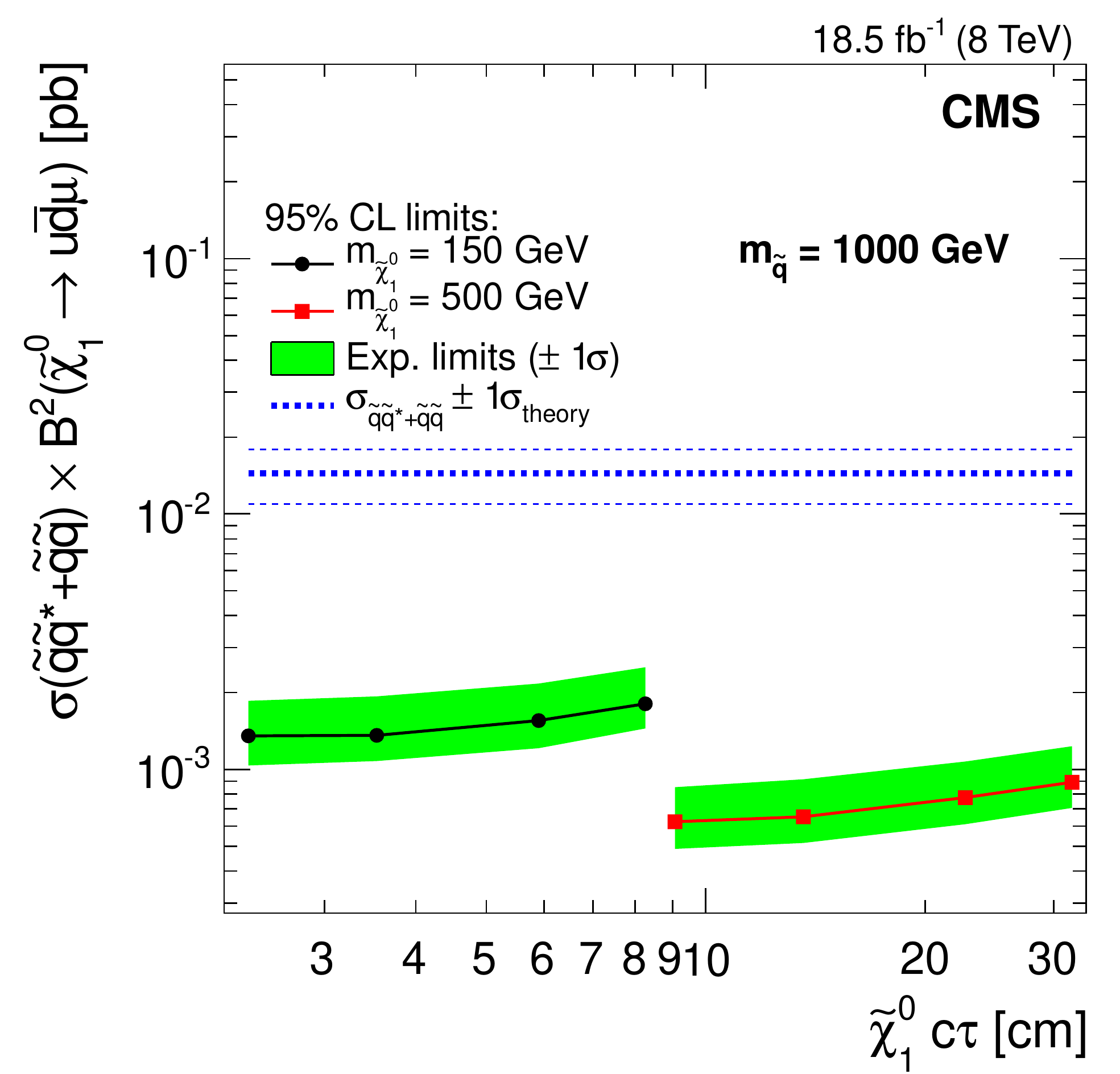}
\includegraphics[width=0.49\textwidth]{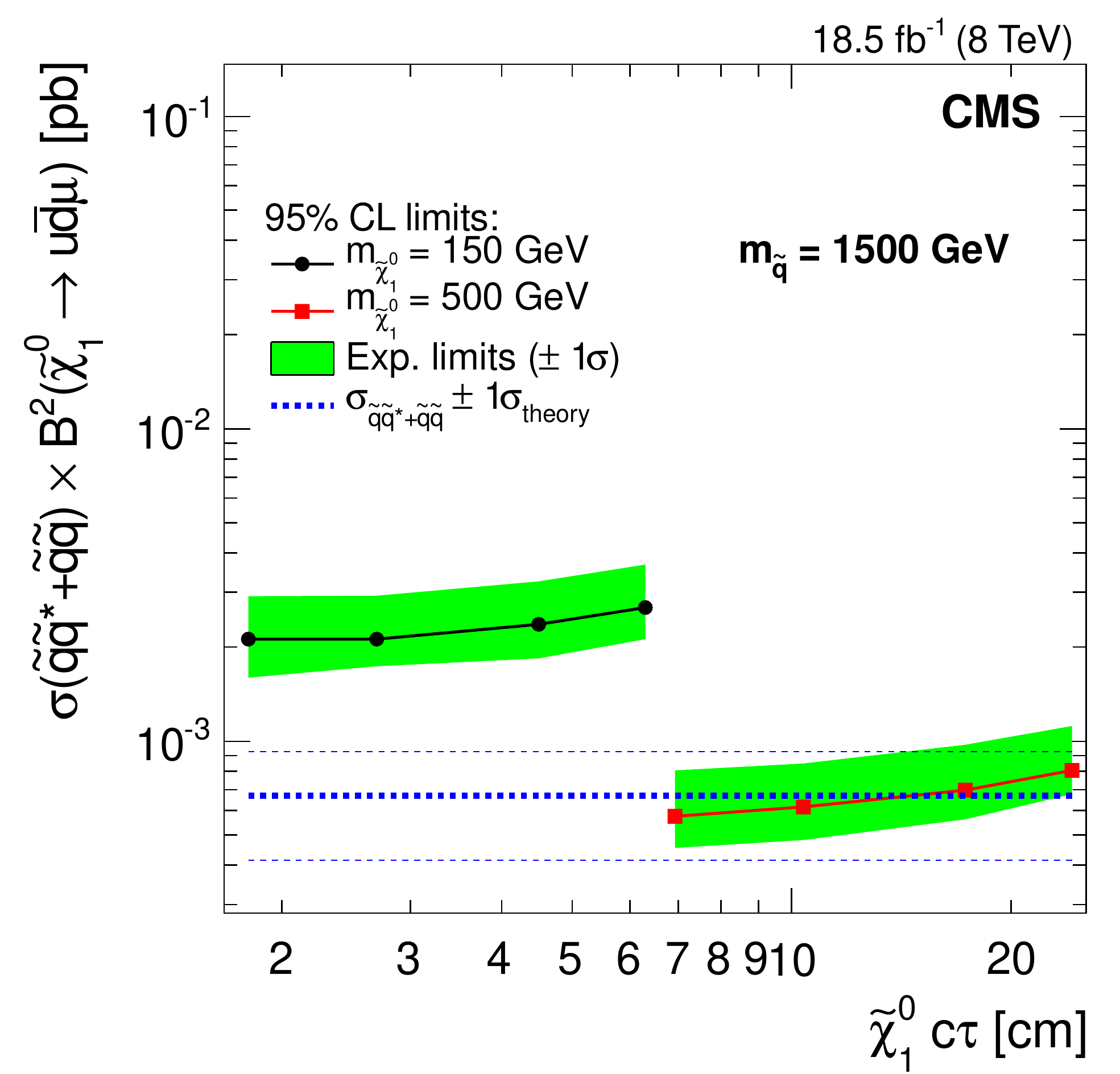}
\caption{The 95\% \CL upper limits on the product of the
cross section to produce a pair of squarks, where each squark decays
to a long-lived neutralino, and the branching fraction squared $\mathcal{B}^{2}$ for neutralino to decay into
a pair of up or down quarks and a muon.
The limits are presented as a function of the neutralino mean proper decay length
separately for each squark/neutralino mass point. For each mass point the theoretical
cross section for \sqasq, and its systematic uncertainty, are shown.
Solid bands show the ${\pm}1\sigma$ range of variation of
the expected 95\% \CL limits.
\label{fig:limitsneu}}
\end{figure*}

In Fig.~\ref{fig:limits} we show the upper limits on the product of the cross section to produce $\Higgs \to 2\X$
and the branching fraction squared $\mathcal{B}^2$ for \X to decay into \qq.
The upper limits
on the squark production cross section (where each squark decays to a neutralino that
decays into a quark-antiquark pair and a muon) are presented in
Fig.~\ref{fig:limitsneu}.
In order to increase the number of tested models,
the lifetime distributions of the signal long-lived particles are reweighted to
different mean values, between $0.4\tau$ and $1.4\tau$, for every
lifetime value $\tau$ and mass combination listed
in Table \ref{tab:sigeff}.
Event weights are computed as the product
 of weights assigned to each long-lived particle in the event.
The reweighted signal reconstruction efficiencies are then used
to compute the expected and observed limits for the additional mean lifetime values.
The upper limits for the neutralino model are compared with
NLO calculations of the squark production
cross section, including next-to-leading-logarithmic (NLL) corrections obtained with the program
\PROSPINO \cite{Beenakker:1996ch,Kulesza:2008jb,Kulesza:2009kq}. The theoretical cross
section for \sqasq is 10, 0.139, 0.014, and 0.00067\unit{pb} for \squark masses of 350, 700, 1000, and 1500\GeV, respectively, assuming a gluino mass of 5\TeV. The cross section uncertainty
band represents the variation of the QCD factorization and renormalization scales, each
up and down by a factor of 2, as well as a variation obtained by
using two different sets of NLO parton distribution
functions (CTEQ6.6 and MSTW2008 \cite{Martin:2009iq}).

When a neutralino decays into a quark-antiquark pair and a muon, all three particles
may be identified as jets by the jet reconstruction algorithm.
The selected dijet candidate can therefore be formed from a quark-quark
or a quark-muon pair. There are up to six displaced dijet pairings per event, two quark-quark pairs
and four quark-muon pairs. Using $\Delta R$ matching between the generator-level particles
 and reconstructed jets, we find that at least 50\% of the
 accepted events have a dijet candidate selected that is associated with a quark-quark pair, for all
squark/neutralino masses.

\section{Summary}
\label{sec:summary}

A search for long-lived particles, produced in proton-proton collisions at $\sqrt{s} = 8\TeV$ and decaying
to quark-antiquark pairs, has been performed.
The observed results are consistent with standard model expectations and are used to derive upper
limits on the product of cross section and branching fraction for a scalar particle
\Higgs in the mass range 200 to 1000\GeV, decaying promptly into a pair of long-lived \X bosons in the mass
range 50 to 350\GeV, which further decay to quark-antiquark pairs. For \X mean proper decay lengths
in the range 0.4 to 200\cm, the upper limits are typically 0.5--200\unit{fb}.
Additionally, the results are interpreted
for the pair-production of long-lived neutralinos
that decay into two quarks and a muon through an R-parity violating coupling.
For pair production of squarks, which promptly decay to
neutralinos that have mean proper decay lengths in the range 2 to 40\cm, the upper limits on the cross section
are typically 0.5--3\unit{fb}.
The above limits are the most stringent on these channels to date.

\begin{acknowledgments}
We congratulate our colleagues in the CERN accelerator departments for the excellent performance of the LHC and thank the technical and administrative staffs at CERN and at other CMS institutes for their contributions to the success of the CMS effort. In addition, we gratefully acknowledge the computing centers and personnel of the Worldwide LHC Computing Grid for delivering so effectively the computing infrastructure essential to our analyses. Finally, we acknowledge the enduring support for the construction and operation of the LHC and the CMS detector provided by the following funding agencies: BMWFW and FWF (Austria); FNRS and FWO (Belgium); CNPq, CAPES, FAPERJ, and FAPESP (Brazil); MES (Bulgaria); CERN; CAS, MoST, and NSFC (China); COLCIENCIAS (Colombia); MSES and CSF (Croatia); RPF (Cyprus); MoER, ERC IUT and ERDF (Estonia); Academy of Finland, MEC, and HIP (Finland); CEA and CNRS/IN2P3 (France); BMBF, DFG, and HGF (Germany); GSRT (Greece); OTKA and NIH (Hungary); DAE and DST (India); IPM (Iran); SFI (Ireland); INFN (Italy); MSIP and NRF (Republic of Korea); LAS (Lithuania); MOE and UM (Malaysia); CINVESTAV, CONACYT, SEP, and UASLP-FAI (Mexico); MBIE (New Zealand); PAEC (Pakistan); MSHE and NSC (Poland); FCT (Portugal); JINR (Dubna); MON, RosAtom, RAS and RFBR (Russia); MESTD (Serbia); SEIDI and CPAN (Spain); Swiss Funding Agencies (Switzerland); MST (Taipei); ThEPCenter, IPST, STAR and NSTDA (Thailand); TUBITAK and TAEK (Turkey); NASU and SFFR (Ukraine); STFC (United Kingdom); DOE and NSF (USA).

Individuals have received support from the Marie-Curie program and the European Research Council and EPLANET (European Union); the Leventis Foundation; the A. P. Sloan Foundation; the Alexander von Humboldt Foundation; the Belgian Federal Science Policy Office; the Fonds pour la Formation \`a la Recherche dans l'Industrie et dans l'Agriculture (FRIA-Belgium); the Agentschap voor Innovatie door Wetenschap en Technologie (IWT-Belgium); the Ministry of Education, Youth and Sports (MEYS) of the Czech Republic; the Council of Science and Industrial Research, India; the HOMING PLUS program of Foundation for Polish Science, cofinanced from European Union, Regional Development Fund; the Compagnia di San Paolo (Torino); the Consorzio per la Fisica (Trieste); MIUR project 20108T4XTM (Italy); the Thalis and Aristeia programmes cofinanced by EU-ESF and the Greek NSRF; and the National Priorities Research Program by Qatar National Research Fund.
\end{acknowledgments}
\bibliography{auto_generated}

\providecommand{\href}[2]{#2}\begingroup\raggedright\begin{thebibliography}{10}%
\makeatletter
\providecommand{\hrefCMSnoop }[0]{\@secondoftwo}%
\makeatother
\providecommand{\doi}{\texttt{doi:}\begingroup \urlstyle{tt}\Url}

\bibitem{Hewett:2004nw}
\hrefCMSnoop {}{J.~L. Hewett, B.~Lillie, M.~Masip, and T.~G. Rizzo,
  ``{Signatures of long-lived gluinos in split supersymmetry}'',} \textit{
  JHEP} \textbf{ 09} (2004) 070,
  \href{http://dx.doi.org/10.1088/1126-6708/2004/09/070}{\doi{10.1088/1126-6708/2004/09/070}},
  \href{http://www.arXiv.org/abs/hep-ph/0408248}{\texttt{
  arXiv:hep-ph/0408248}}.

\bibitem{Barbier:2004ez}
R.~Barbieri\hrefCMSnoop {}{ {et~al.}, ``{R}-Parity violating supersymmetry'',}
  \textit{ Phys. Rept.} \textbf{ 420} (2005) 1,
  \href{http://dx.doi.org/10.1016/j.physrep.2005.08.006}{\doi{10.1016/j.physrep.2005.08.006}},
\href{http://www.arXiv.org/abs/hep-ph/0406039}{\texttt{ arXiv:hep-ph/0406039}}.

\bibitem{Han:2007ae}
\hrefCMSnoop {}{T.~Han, Z.~Si, K.~M. Zurek, and M.~J. Strassler,
  ``{Phenomenology of hidden valleys at hadron colliders}'',} \textit{ JHEP}
  \textbf{ 07} (2008) 008,
  \href{http://dx.doi.org/10.1088/1126-6708/2008/07/008}{\doi{10.1088/1126-6708/2008/07/008}},
  \href{http://www.arXiv.org/abs/0712.2041}{\texttt{ arXiv:0712.2041}}.

\bibitem{Basso:2008iv}
\hrefCMSnoop {}{L.~Basso, A.~Belyaev, S.~Moretti, and C.~H.
  Shepherd-Themistocleous, ``{Phenomenology of the minimal B-L extension of the
  standard model: Z' and neutrinos}'',} \textit{ Phys. Rev. D} \textbf{ 80}
  (2009) 055030,
  \href{http://dx.doi.org/10.1103/PhysRevD.80.055030}{\doi{10.1103/PhysRevD.80.055030}},
  \href{http://www.arXiv.org/abs/0812.4313}{\texttt{ arXiv:0812.4313}}.

\bibitem{Strassler:2006ri}
\hrefCMSnoop {}{M.~J. Strassler and K.~M. Zurek, ``{Discovering the Higgs
  through highly-displaced vertices}'',} \textit{ Phys. Lett. B} \textbf{ 661}
  (2008) 263,
  \href{http://dx.doi.org/10.1016/j.physletb.2008.02.008}{\doi{10.1016/j.physletb.2008.02.008}},
  \href{http://www.arXiv.org/abs/hep-ph/0605193}{\texttt{
  arXiv:hep-ph/0605193}}.

\bibitem{Aaltonen:2011rja}
\hrefCMSnoop {}{{CDF} Collaboration, ``{Search for heavy metastable particles
  decaying to jet pairs in $\text{p}\overline{\text p}$ collisions at $\sqrt{s}
  = 1.96$ TeV}'',} \textit{ Phys. Rev. D} \textbf{ 85} (2012) 012007,
  \href{http://dx.doi.org/10.1103/PhysRevD.85.012007}{\doi{10.1103/PhysRevD.85.012007}},
\href{http://www.arXiv.org/abs/1109.3136}{\texttt{ arXiv:1109.3136}}.

\bibitem{Abazov:2009ik}
\hrefCMSnoop {}{{D0} Collaboration, ``{Search for resonant pair production of
  neutral long-lived particles decaying to \bbbar in $\text{p} \overline{\text
  p}$ collisions at $\sqrt{s}= 1.96$ TeV}'',} \textit{ Phys. Rev. Lett.}
  \textbf{ 103} (2009) 071801,
  \href{http://dx.doi.org/10.1103/PhysRevLett.103.071801}{\doi{10.1103/PhysRevLett.103.071801}},
\href{http://www.arXiv.org/abs/0906.1787}{\texttt{ arXiv:0906.1787}}.

\bibitem{ATLAS:2012av}
\hrefCMSnoop {}{{ATLAS} Collaboration, ``{Search for a light Higgs boson
  decaying to long-lived weakly-interacting particles in proton-proton
  collisions at $\sqrt{s}=7$ TeV with the ATLAS detector}'',} \textit{ Phys.
  Rev. Lett.} \textbf{ 108} (2012) 251801,
  \href{http://dx.doi.org/10.1103/PhysRevLett.108.251801}{\doi{10.1103/PhysRevLett.108.251801}},
\href{http://www.arXiv.org/abs/1203.1303}{\texttt{ arXiv:1203.1303}}.

\bibitem{Aad:2011zb}
\hrefCMSnoop {}{{ATLAS} Collaboration, ``{Search for displaced vertices arising
  from decays of new heavy particles in 7 TeV pp collisions at ATLAS}'',}
  \textit{ Phys. Lett. B} \textbf{ 707} (2012) 478,
  \href{http://dx.doi.org/10.1016/j.physletb.2011.12.057}{\doi{10.1016/j.physletb.2011.12.057}},
\href{http://www.arXiv.org/abs/1109.2242}{\texttt{ arXiv:1109.2242}}.

\bibitem{Chatrchyan:2013oca}
\hrefCMSnoop {}{{CMS} Collaboration, ``{Searches for long-lived charged
  particles in pp collisions at $\sqrt{s}$ = 7 and 8 TeV}'',} \textit{ JHEP}
  \textbf{ 07} (2013) 122,
  \href{http://dx.doi.org/10.1007/JHEP07(2013)122}{\doi{10.1007/JHEP07(2013)122}},
\href{http://www.arXiv.org/abs/1305.0491}{\texttt{ arXiv:1305.0491}}.

\bibitem{Chatrchyan:2012ir}
\hrefCMSnoop {}{{CMS} Collaboration, ``{Search for new physics with long-lived
  particles decaying to photons and missing energy in pp collisions at
  $\sqrt{s}=7$ TeV}'',} \textit{ JHEP} \textbf{ 11} (2012) 172,
  \href{http://dx.doi.org/10.1007/JHEP11(2012)172}{\doi{10.1007/JHEP11(2012)172}},
\href{http://www.arXiv.org/abs/1207.0627}{\texttt{ arXiv:1207.0627}}.

\bibitem{Chatrchyan:2012jna}
\hrefCMSnoop {}{{CMS} Collaboration, ``{Search in leptonic channels for heavy
  resonances decaying to long-lived neutral particles}'',} \textit{ JHEP}
  \textbf{ 02} (2013) 085,
  \href{http://dx.doi.org/10.1007/JHEP02(2013)085}{\doi{10.1007/JHEP02(2013)085}},
\href{http://www.arXiv.org/abs/1211.2472}{\texttt{ arXiv:1211.2472}}.

\bibitem{Chatrchyan:2012dxa}
\hrefCMSnoop {}{{CMS} Collaboration, ``{Search for stopped long-lived particles
  produced in pp collisions at $\sqrt{s}=7$ TeV}'',} \textit{ JHEP} \textbf{
  08} (2012) 026,
  \href{http://dx.doi.org/10.1007/JHEP08(2012)026}{\doi{10.1007/JHEP08(2012)026}},
\href{http://www.arXiv.org/abs/1207.0106}{\texttt{ arXiv:1207.0106}}.

\bibitem{Chatrchyan:2008zzk}
\hrefCMSnoop {}{{CMS} Collaboration, ``The {CMS} experiment at the {CERN}
  {LHC}'',} \textit{ JINST} \textbf{ 03} (2008) S08004,
  \href{http://dx.doi.org/10.1088/1748-0221/3/08/S08004}{\doi{10.1088/1748-0221/3/08/S08004}}.

\bibitem{Chatrchyan:2014fea}
\hrefCMSnoop {}{{CMS} Collaboration, ``{Description and performance of track
  and primary-vertex reconstruction with the CMS tracker}'',} \textit{ JINST}
  \textbf{ 9} (2014) P10009,
  \href{http://dx.doi.org/10.1088/1748-0221/9/10/P10009}{\doi{10.1088/1748-0221/9/10/P10009}},
\href{http://www.arXiv.org/abs/1405.6569}{\texttt{ arXiv:1405.6569}}.

\bibitem{CMS-PAS-PFT-09-001}
\href {http://cdsweb.cern.ch/record/1194487}{{CMS} Collaboration,
  ``Particle--Flow Event Reconstruction in {CMS} and Performance for Jets,
  Taus, and {\MET}'',} CMS Physics Analysis Summary CMS-PAS-PFT-09-001, 2009.

\bibitem{CMS-PAS-PFT-10-001}
\href {http://cdsweb.cern.ch/record/1247373}{{CMS} Collaboration,
  ``Commissioning of the particle-flow event reconstruction with the first
  {LHC} collisions recorded in the {CMS} detector'',} CMS Physics Analysis
  Summary CMS-PAS-PFT-10-001, 2010.

\bibitem{Cacciari:2008gp}
\hrefCMSnoop {}{M.~Cacciari, G.~P. Salam, and G.~Soyez, ``The anti-$k_t$ jet
  clustering algorithm'',} \textit{ JHEP} \textbf{ 04} (2008) 063,
  \href{http://dx.doi.org/10.1088/1126-6708/2008/04/063}{\doi{10.1088/1126-6708/2008/04/063}},
  \href{http://www.arXiv.org/abs/0802.1189}{\texttt{ arXiv:0802.1189}}.

\bibitem{Chatrchyan:2011ds}
\hrefCMSnoop {}{{CMS} Collaboration, ``{Determination of Jet Energy Calibration
  and Transverse Momentum Resolution in CMS}'',} \textit{ JINST} \textbf{ 6}
  (2011) P11002,
  \href{http://dx.doi.org/10.1088/1748-0221/6/11/P11002}{\doi{10.1088/1748-0221/6/11/P11002}},
\href{http://www.arXiv.org/abs/1107.4277}{\texttt{ arXiv:1107.4277}}.

\bibitem{Cacciari:2007fd}
\hrefCMSnoop {}{M.~Cacciari and G.~P. Salam, ``{Pileup subtraction using jet
  areas}'',} \textit{ Phys. Lett. B} \textbf{ 659} (2008) 119,
  \href{http://dx.doi.org/10.1016/j.physletb.2007.09.077}{\doi{10.1016/j.physletb.2007.09.077}},
\href{http://www.arXiv.org/abs/0707.1378}{\texttt{ arXiv:0707.1378}}.

\bibitem{CMS-PAS-LUM-13-001}
\href {http://cdsweb.cern.ch/record/1598864}{{CMS} Collaboration, ``CMS
  Luminosity Based on Pixel Cluster Counting - Summer 2013 Update'',} CMS
  Physics Analysis Summary CMS-PAS-LUM-13-001, 2013.

\bibitem{PYTHIA}
\hrefCMSnoop {}{T.~Sj{\"o}strand, S.~Mrenna, and P.~Z. Skands, ``{PYTHIA} 6.4
  physics and manual'',} \textit{ JHEP} \textbf{ 05} (2006) 026,
  \href{http://dx.doi.org/10.1088/1126-6708/2006/05/026}{\doi{10.1088/1126-6708/2006/05/026}},
\href{http://www.arXiv.org/abs/hep-ph/0603175}{\texttt{ arXiv:hep-ph/0603175}}.

\bibitem{Pumplin:2002vw}
J.~Pumplin\hrefCMSnoop {}{ {et~al.}, ``{New generation of parton distributions
  with uncertainties from global QCD analysis}'',} \textit{ JHEP} \textbf{ 07}
  (2002) 012,
  \href{http://dx.doi.org/10.1088/1126-6708/2002/07/012}{\doi{10.1088/1126-6708/2002/07/012}},
  \href{http://www.arXiv.org/abs/hep-ph/0201195}{\texttt{
  arXiv:hep-ph/0201195}}.

\bibitem{GEANT4}
\hrefCMSnoop {}{{GEANT4} Collaboration, ``{GEANT4}---a simulation toolkit'',}
  \textit{ Nucl. Instrum. Meth. A} \textbf{ 506} (2003) 250,
\href{http://dx.doi.org/10.1016/S0168-9002(03)01368-8}{\doi{10.1016/S0168-9002(03)01368-8}}.

\bibitem{Waltenberger:1166320}
\href {http://cds.cern.ch/record/1166320}{W.~Waltenberger, ``{Adaptive Vertex
  Reconstruction}'',} CMS NOTE 2008-033, 2008.

\bibitem{opac-b1127878}
T.~J. Hastie, R.~J. Tibshirani, and J.~H. Friedman, ``The elements of
  statistical learning : data mining, inference, and prediction''.
\newblock Springer series in statistics. Springer, New York, 2009.

\bibitem{Kirschenmann:2012toa}
\hrefCMSnoop {}{{CMS} Collaboration, ``{Determination of the jet energy scale
  in CMS}'',} \textit{ J. Phys. Conf. Ser.} \textbf{ 404} (2012) 012013,
\href{http://dx.doi.org/10.1088/1742-6596/404/1/012013}{\doi{10.1088/1742-6596/404/1/012013}}.

\bibitem{Beringer:1900zz}
\hrefCMSnoop {}{{Particle Data Group}, J.~Beringer {et~al.}, ``{Review of
  Particle Physics}'',} \textit{ Phys. Rev. D} \textbf{ 86} (2012) 010001,
\href{http://dx.doi.org/10.1103/PhysRevD.86.010001}{\doi{10.1103/PhysRevD.86.010001}}.

\bibitem{Khachatryan:2011tm}
\hrefCMSnoop {}{{CMS} Collaboration, ``Strange particle production in pp
  collisions at $\sqrt{s}$ = 0.9 and 7 {TeV}'',} \textit{ JHEP} \textbf{ 05}
  (2011) 064,
  \href{http://dx.doi.org/10.1007/JHEP05(2011)064}{\doi{10.1007/JHEP05(2011)064}},
\href{http://www.arXiv.org/abs/1102.4282}{\texttt{ arXiv:1102.4282}}.

\bibitem{Nason:2004rx}
\hrefCMSnoop {}{P.~Nason, ``{A New method for combining NLO QCD with shower
  Monte Carlo algorithms}'',} \textit{ JHEP} \textbf{ 11} (2004) 040,
  \href{http://dx.doi.org/10.1088/1126-6708/2004/11/040}{\doi{10.1088/1126-6708/2004/11/040}},
\href{http://www.arXiv.org/abs/hep-ph/0409146}{\texttt{ arXiv:hep-ph/0409146}}.

\bibitem{Frixione:2007vw}
\hrefCMSnoop {}{S.~Frixione, P.~Nason, and C.~Oleari, ``{Matching NLO QCD
  computations with Parton Shower simulations: the POWHEG method}'',} \textit{
  JHEP} \textbf{ 11} (2007) 070,
  \href{http://dx.doi.org/10.1088/1126-6708/2007/11/070}{\doi{10.1088/1126-6708/2007/11/070}},
\href{http://www.arXiv.org/abs/0709.2092}{\texttt{ arXiv:0709.2092}}.

\bibitem{Alioli:2010xd}
\hrefCMSnoop {}{S.~Alioli, P.~Nason, C.~Oleari, and E.~Re, ``{A general
  framework for implementing NLO calculations in shower Monte Carlo programs:
  the POWHEG BOX}'',} \textit{ JHEP} \textbf{ 06} (2010) 043,
  \href{http://dx.doi.org/10.1007/JHEP06(2010)043}{\doi{10.1007/JHEP06(2010)043}},
\href{http://www.arXiv.org/abs/1002.2581}{\texttt{ arXiv:1002.2581}}.

\bibitem{Read:2002hq}
\hrefCMSnoop {}{A.~L. Read, ``Presentation of search results: the {$CL_s$}
  technique'',} \textit{ J. Phys. G} \textbf{ 28} (2002) 2693,
\href{http://dx.doi.org/10.1088/0954-3899/28/10/313}{\doi{10.1088/0954-3899/28/10/313}}.

\bibitem{Junk:1999kv}
\hrefCMSnoop {}{T.~Junk, ``{Confidence level computation for combining searches
  with small statistics}'',} \textit{ Nucl. Instrum. Meth. A} \textbf{ 434}
  (1999) 435,
  \href{http://dx.doi.org/10.1016/S0168-9002(99)00498-2}{\doi{10.1016/S0168-9002(99)00498-2}},
\href{http://www.arXiv.org/abs/hep-ex/9902006}{\texttt{ arXiv:hep-ex/9902006}}.

\bibitem{Beenakker:1996ch}
\hrefCMSnoop {}{W.~Beenakker, R.~H{\"o}pker, M.~Spira, and P.~M. Zerwas,
  ``{Squark and gluino production at hadron colliders}'',} \textit{ Nucl. Phys.
  B} \textbf{ 492} (1997) 51,
  \href{http://dx.doi.org/10.1016/S0550-3213(97)80027-2}{\doi{10.1016/S0550-3213(97)80027-2}},
\href{http://www.arXiv.org/abs/hep-ph/9610490}{\texttt{ arXiv:hep-ph/9610490}}.

\bibitem{Kulesza:2008jb}
\hrefCMSnoop {}{A.~Kulesza and L.~Motyka, ``{Threshold resummation for
  squark-antisquark and gluino-pair production at the LHC}'',} \textit{ Phys.
  Rev. Lett.} \textbf{ 102} (2009) 111802,
  \href{http://dx.doi.org/10.1103/PhysRevLett.102.111802}{\doi{10.1103/PhysRevLett.102.111802}},
\href{http://www.arXiv.org/abs/0807.2405}{\texttt{ arXiv:0807.2405}}.

\bibitem{Kulesza:2009kq}
\hrefCMSnoop {}{A.~Kulesza and L.~Motyka, ``{Soft gluon resummation for the
  production of gluino-gluino and squark-antisquark pairs at the LHC}'',}
  \textit{ Phys. Rev. D} \textbf{ 80} (2009) 095004,
  \href{http://dx.doi.org/10.1103/PhysRevD.80.095004}{\doi{10.1103/PhysRevD.80.095004}},
\href{http://www.arXiv.org/abs/0905.4749}{\texttt{ arXiv:0905.4749}}.

\bibitem{Martin:2009iq}
\hrefCMSnoop {}{A.~D. Martin, W.~J. Stirling, R.~S. Thorne, and G.~Watt,
  ``Parton distributions for the {LHC}'',} \textit{ Eur. Phys. J. C} \textbf{
  63} (2009) 189,
  \href{http://dx.doi.org/10.1140/epjc/s10052-009-1072-5}{\doi{10.1140/epjc/s10052-009-1072-5}},
\href{http://www.arXiv.org/abs/0901.0002}{\texttt{ arXiv:0901.0002}}.

\end{thebibliography}\endgroup

\cleardoublepage \appendix\section{The CMS Collaboration \label{app:collab}}\begin{sloppypar}\hyphenpenalty=5000\widowpenalty=500\clubpenalty=5000\textbf{Yerevan Physics Institute,  Yerevan,  Armenia}\\*[0pt]
V.~Khachatryan, A.M.~Sirunyan, A.~Tumasyan
\vskip\cmsinstskip
\textbf{Institut f\"{u}r Hochenergiephysik der OeAW,  Wien,  Austria}\\*[0pt]
W.~Adam, T.~Bergauer, M.~Dragicevic, J.~Er\"{o}, C.~Fabjan\cmsAuthorMark{1}, M.~Friedl, R.~Fr\"{u}hwirth\cmsAuthorMark{1}, V.M.~Ghete, C.~Hartl, N.~H\"{o}rmann, J.~Hrubec, M.~Jeitler\cmsAuthorMark{1}, W.~Kiesenhofer, V.~Kn\"{u}nz, M.~Krammer\cmsAuthorMark{1}, I.~Kr\"{a}tschmer, D.~Liko, I.~Mikulec, D.~Rabady\cmsAuthorMark{2}, B.~Rahbaran, H.~Rohringer, R.~Sch\"{o}fbeck, J.~Strauss, A.~Taurok, W.~Treberer-Treberspurg, W.~Waltenberger, C.-E.~Wulz\cmsAuthorMark{1}
% ...
\vskip\cmsinstskip
% ...
1:~~Also at Vienna University of Technology, Vienna, Austria\\
2:~~Also at CERN, European Organization for Nuclear Research, Geneva, Switzerland\\
% ...

\end{sloppypar}
\end{document}